\documentclass[12pt,a4paper]{article}
\usepackage{amsmath}
\usepackage{amssymb}
\usepackage{enumerate}

\newcommand{\be}{\begin{equation}}
\newcommand{\ee}{\end{equation}}
\newcommand{\bea}{\begin{eqnarray}}
\newcommand{\eea}{\end{eqnarray}}

\makeatletter
\def\section{\@startsection{section}{1}
{\z@}{-3.5ex plus -1ex minus -.2ex}{2.3ex plus .2ex}{\normalsize \bf}}
\def\subsection{\@startsection{subsection}{2}
{\z@}{-3.25ex plus -1ex minus -.2ex}{1.5ex plus .2ex}{\normalsize \sl}}
\def\subsubsection{\@startsection{subsubsection}{3}
{\z@}{-3.25ex plus -1ex minus -.2ex}{1.5ex plus .2ex}{\normalsize}}

\def\evenmark{\bf entropy}
\def\oddmarkA{{\sl Entropy} {\bf 2005}, {\sl 7[4]}, 253-299}
\def\oddmarkC{}
\def\oddmarkD{{\Huge \bf Entropy}}
\def\oddmarkE{\normalsize {\bf ISSN 1099-4300}}
\def\oddmarkF{{www.mdpi.org/entropy/}}

\makeatletter
\def\@evenhead{
    \vbox{\hbox to\hsize{\bf \thepage \hfill \sl \evenmark}}
}
\def\@oddhead{
    \vbox{
        \hbox to\hsize{\oddmarkA \hfill \oddmarkC}
        \hbox to\hsize{\hfill \oddmarkD}
        \vspace{.03in}
        \hbox to\hsize{\hfill \oddmarkE}
        \vspace{.03in}
        \hbox to\hsize{\hfill \oddmarkF}
    }
}
\makeatother

\def\papercategory{
Review }

\def\thedoctitle{\bf
Differential entropy and time }

\def\theauthorname{
Piotr Garbaczewski }

\def\authoraddress{Institute of Physics,  University  of Zielona
G\'{o}ra, ul. Szafrana 4a,  65-516 Zielona G\'{o}ra, Poland
\\ E-mail: p.garbaczewski@if.uz.zgora.pl
}

\def\thereceivedhistory{
Received: 22 August 2005 / Accepted: 17 October 2005 / Published:
18 October 2005}

\begin{document}
\parindent 0cm

{ \noindent {\\ \\} }

{ {\sl \papercategory} } \\

{ \LARGE \noindent \thedoctitle } \\

{ \noindent {\bf {\theauthorname}} } \\

{ \normalsize \noindent {\authoraddress} } \\

{ \noindent {\sl \thereceivedhistory} } \\

\vspace{2pt} \hbox to \hsize{\hrulefill}
\vspace{.1in}

\noindent

{\bf Abstract:}  We give a detailed analysis  of  the   Gibbs-type
entropy notion  and its  dynamical behavior in  case of
time-dependent continuous probability distributions of varied
origins: related to classical  and quantum systems. The
purpose-dependent usage of conditional Kullback-Leibler and Gibbs
(Shannon) entropies is explained in case of non-equilibrium
Smoluchowski processes. A very different temporal behavior of
Gibbs and Kullback entropies  is confronted. A specific conceptual
niche is addressed, where quantum von Neumann, classical
Kullback-Leibler and  Gibbs entropies can be consistently
introduced as information measures for the same physical system.
If the dynamics of probability densities is driven by the
Schr\"{o}dinger picture wave-packet evolution, Gibbs-type  and
related Fisher information functionals appear to quantify
nontrivial power transfer processes in the mean.  This observation
is found to extend to classical dissipative processes and supports
the view that the Shannon entropy dynamics provides an insight
into physically relevant non-equilibrium phenomena, which are
inaccessible in terms of the Kullback-Leibler entropy and
typically ignored in the literature.
\\

\\



\vspace{2pt} \hbox to \hsize{\hrulefill}

\newpage

\def\oddmarkC{\thepage}
\def\oddmarkD{}
\def\oddmarkE{}
\def\oddmarkF{}

\section{Introduction}

Among numerous  manifestations of the concept of entropy in
physics and mathematics, the  information-theory based entropy
methods have been  devised to investigate the large time behavior
of solutions of  various (mostly dissipative)  partial
differential equations. Shannon, Kullback and von Neumann
entropies are typical information  theory   tools,    designed  to
quantify the information content and possibly information loss for
various classical and quantum  systems  \it  in a specified
state.\rm

  For quantum systems the von Neumann entropy vanishes on
 pure states, hence   one presumes to have   a complete information about
the state of a system.  On the other  hand, for pure states the
Gibbs-type  (Shannon, e.g. differential) entropy   gives access to
another information-theory  level, associated with a probability
distribution inferred from a $L^2(R^n)$  wave packet.

 Although  Shannon or Kullback  entropies  are  interpreted as   information
  measures,   it is   quite natural to think of entropy as  a
  measure of uncertainty. In view of the profound role played by the
  Shannon entropy  in the formulation of entropic indeterminacy relations
  \cite{alicki,petz},  the term \it  information, \rm   in the present paper
  is mostly  used  in the technical sense, meaning the inverse of \it
   uncertainty.\rm

In  physics,  the notion of entropy is  typically regarded as a
measure of the degree of randomness and the tendency (trends)
  of physical  systems to become less and less organized.
Throughout the paper  we shall attribute  a  more concrete meaning
to the term \it  organization,  \rm   both in the classical  and
quantum  contexts.  Namely, we shall pay special attention to
quantifying, in terms of suitable entropy functionals, the degree
of the probability distribution  (de)localization, and the
dynamical behavior  of this specific - localization uncertainty -
feature of a physical system.

\subsection{Notions of entropy}

Notions of entropy, information and  uncertainty are intertwined
and cannot be sharply differentiated.  While  entropy  and
uncertainty are - to some extent synonymous - measures of
ignorance (lack of information, uncertainty), the complementary
notion of information basically quantifies the \it  ability \rm of
observers to make reliable predictions about the system,
\cite{shannon,sobczyk,yaglom}:
 the more aware  one is  about chances of a concrete outcome, the lower
 is the uncertainty of this outcome.

 Normally,  the growth of uncertainty
is identified with an increase of the entropy which in turn  is
interpreted as an information loss. Consult e.g. standard
formulations of the celebrated Boltzmann H-theorem.

Following Ref.~\cite{wehrl} let us recall that   entropy - be it
thermodynamical (Gibbs-Boltzmann), dynamical,   von Neumann,
Wehrl,  Shannon, Renyi, Tsallis   or any other conceivable
candidate - has an exceptional status among physical quantities.
As a \it derived \rm quantity it does not show up in  any
fundamental equation of motion.  Generically, there is no a priori
preferred  notion of entropy (perhaps, except  for the
thermodynamical Clausius  case) in physical applications  and its
specific choice  appears to be purpose-dependent.

As an obvious remnant of the standard  thermodynamical reasoning,
one expects  entropy to be   a  state  function  of the system
(thermodynamical notions of equilibrium or   near-equilibrium are
implicit).  This  \it  state \rm  connotation is a
 source of ambiguities, since inequivalent notions  of the system  state
 are used in the  description of physical systems, be them  classical, thermodynamical and
 quantum. Not to mention rather specialized meaning of \it  state \rm  employed in
 the standard  information theory, \cite{shannon,bril,yaglom}.

A primitive information-theory system  is  simply a   \it bit \rm
whose two admissible  states are binary digits $1$ and $0$.
 Its quantum equivalent is a \it qubit \rm whose admissible states are
  vectors in  a two-dimensional Hilbert space, hence an infinity of  pure states of
   a two-level quantum system.

The   information theory  framework, if   extended to more
complicated systems,   employs   a plethora of  notions of state
\cite{shannon,yaglom}.
 As very special cases we  may mention  a phase-space point
  as  the  determinative of the state of a classical dynamical  system,
  or   the   macroscopic  notion of a thermodynamical state in its classical and quantum versions,
  \cite{wehrl}.

 A  symbolic mathematical representation of quantum states in
terms of wave vectors and/or density operators is expected to
provide  an experimentally verifiable "information" about the
system. To obtain a catalogue of the corresponding statistical
predictions, an a priori choice of suitable observables (and thus
measurement procedures)  is necessary. Then, a  casual
interpretation of entropy as a measure of one's \it uncertainty
\rm about measurable properties of a system in a prescribed
quantum state may acquire an unambiguous meaning.

 When  adopting   the state notion   to the Hilbert space language of quantum theory,
 we realize that   normalized   wave functions  and  density operators, which are traditionally
  supposed to determine the quantum state,  allow to
 extend the  notion of entropy  to  certain functionals
   of  the  state  of the quantum system.  The von Neumann entropy
\begin{equation}
{\cal{S}}(\hat{\rho })= - k_B \, Tr (\hat{\rho } \ln \hat{\rho })
\end{equation}
 of  a quantum state (e.g. the  density operator $\hat{\rho }$), though  often infinite, is typically
related to  the degree  of departure   from purity (e.g. the
"mixedness" level)  of the state and is particularly  useful while
quantifying  measurements performed upon finite quantum systems.

 For a given density operator $\hat{\rho }$,  von Neumann entropy is commonly
   accepted  as a reliable measure of the information
content (about the  departure from purity),  to be  experimentally
extracted  from of a quantum system in a given state. Only under
very specific circumstances,  like e.g. in an optimal "quantum
experiment" \cite{brukner,mana} which refers to the diagonal
density operator (with  $p_i$, $1\leq i\leq N$ being its
eigenvalues),
 the information gain  can be described in terms of
 both von Neumann's and  the standard  Shannon measure of information:
\begin{equation}
   -Tr (\hat{\rho } \ln \hat{\rho }) =
 - \sum_i p_i \ln p_i \, .
 \end{equation}

Since  von Neumann entropy is invariant under unitary
transformations, the result exhibits an invariance under the
change of the Hilbert space basis and the  conservation in time
for a closed system (when there is no information/energy exchange
with the environment). Thus,  Schr\"{o}dinger dynamics has no
impact on the von Neumann  encoding of information, see e.g. also
\cite{jaynes,stotland} for a related discussion.

Pure states have vanishing von Neumann entropy
(${\cal{S}}(\hat{\rho })=0$  "for the pure states and only for
them", \cite{wehrl}) and are normally considered as irrelevant
from the quantum information theory perspective, since "one has
complete information" \cite{wehrl} about such states. One may even
say that a pure state is an unjustified over-idealization, since
otherwise it would constitute e.g.  a completely measured  state
of a system in an infinite Hilbert space, \cite{partovi}. A
colloquial  interpretation of this situation is:   since  the wave
function  provides a complete description  of a quantum system,
surely  we have  no uncertainty about this quantum system,
\cite{adami}.

Note that as a side comment we find in Ref.~\cite{partovi} a minor
excuse: this idealization, often employed for position-momentum
degrees of freedom, is usually an adequate
 approximation,   to  be read as an answer to an  objection of Ref.~\cite{deutsch}:
  "although continuous observables such as  the position are familiar enough,
they are really unphysical  idealizations", c.f. in this
connection \cite{karw} for an alternative view.

On the other hand,    the  classic  Shannon entropy   is known to
be  a  natural  measure of the amount of uncertainty related to
measurements  for pairs
 of observables, discrete
 and continuous on an equal footing,  when a quantum system actually  is  in a  \it  pure \rm  state.
  Hence, a properly posed question reveals
obvious uncertainties where at the first  glance  we have no
uncertainty. The related  entropic uncertainty relations  for
finite and infinite quantum systems have received  due attention
in the  literature,  in addition to  direct investigations of the
 configuration space entropic uncertainty/information  measure of
 $L^2(R^n)$  wave packets, \cite{partovi}, \cite{hirschman} -\cite{balian}.

The commonly used in the literature  notions of  Shannon and von
Neumann entropy, although
  coinciding in some cases,  definitely refer to  \it  different \rm  categories of predictions
and information measures for  physical  systems. In contrast to
the  exclusively   quantum concept of
 von Neumann entropy,  the   Shannon entropy - quite   apart from its purely  classical
 provenance - appears to  capture a number of properties of quantum systems which cannot be
 detected  nor described  by means of the  von  Neumann entropy.

 Obviously, there is no use of Shannon entropy if one is interested in verifying
 for mixed  quantum states,   how much actually  a given  state is mixed.
 On the other hand, von Neumann entropy
 appears to be useless in the  analysis of $L^2(R)$ wave    packets and their dynamical
  manifestations  (time-dependent analysis) which are currently in the reach of   experimental
techniques,  \cite{zeilinger,rauch}.
  It is enough to   invoke pure quantum states  in $L^2(R^n)$  and standard
  position-momentum observables  which, quite apart from a hasty
  criticism \cite{deutsch},  still stand for a valid canonical quantization
  cornerstone  of  quantum theory, \cite{karw}.

   Those somewhat underestimated facts seem to  underlie statements about  an inadequacy of
   Shannon entropy in the quantum context, \cite{brukner}, while an equally  valid statement
   is  that the  von Neumann entropy happens to be inadequate. The solution of the dilemma
    lies  in specifying the \it purpose, \rm  see  also \cite{mana}.

\subsection{Differential entropy}

We are  primarily  interested in the \it information content \rm
of  pure  quantum states   in $L^2(R^n)$,  and thus pursue  the
following  (albeit scope-limited, c.f. \cite{zeilinger,rauch} for
experimental justifications)  view: an isolated system  is
represented in quantum mechanics by a state vector  that conveys
statistical \it predictions  \rm for possible  measurement
outcomes.

Consequently, it is  the state vector  which  we regard as  an
information (alternatively,  predictions and uncertainty) resource
  and therefore  questions like, \cite{caves}: how much
information in the state vector or  information about what, may be
considered    meaningful. Let us emphasize  that  we do not
attempt   to  define an  information content of a physical system
as a whole, but rather  we wish  to  set appropriate measures of
uncertainty and information   for  concrete   pure \it states \rm
of a  quantum  system.

A  particular quantum  state  should not be misinterpreted to
provide  a complete  description of the corresponding physical
system itself, \cite{newton}. In fact, when we declare any
Schr\"{o}dinger's  $\psi $  as  the state of a quantum system, we
effectively make a  statement about the probabilities of obtaining
certain results upon measurement of suitable observables, hence we
refer to a definite experimental setup. Therefore, to change  or
influence this state is not quite the same as changing  or
influencing the   system.

Our, still vague notion of  \it  information, \rm   does  not
refer to qubits since
  we shall basically operate in an infinite dimensional Hilbert space. This does
  not prohibit a consistent use of   information-theory concepts, since   an
  analytic information content of a  quantum state vector, in the least
  reduced to a properly handled plane wave,  is not merely an
  abstraction and can    be  dug  out in  realistic
  experiments, including those specific to time-dependent quantum
  mechanics, \cite{rauch,zeilinger}.  On the way one  may   verify   a compliance
  with quantum theory
of   a number of  well defined properties  of the quantum system
for which:  the only features  known before an experiment is
performed  are probabilities of various events to occur,
\cite{brukner}.

In the case of a quantum mechanical  position probability density,
its  analytic form is assumed to  arise in conjunction  with
solutions of the  Schr\"{o}dinger equation. Then, we need to
generalize the   original   Shannon's entropy for a discrete set
of probabilities   to  the entropy of a continuous distribution
with the density distribution function \cite{shannon}, which  is
also  named the differential entropy, \cite{cover,sobczyk}.

Most of our further discussion will be set  in  a specific context
of quantum position-momentum   information/uncertainty   measures,
where  the classical form of   Shannon differential  entropy
\cite{shannon} has  been used for years in the  formulation of
entropic versions of  Heisenberg-type  indeterminacy relations,
\cite{hirschman,beckner,mycielski,uffink}.

The entropic form of indeterminacy relations,   enters  the stage
through  the  Fourier analysis of  $L^2(R^n)$  wave packets,  in
conjunction   with the Born statistical  interpretation, hence
with $\psi $-induced probability measures in position and momentum
space, \cite{hirschman,beckner}. The  experimental connotations
pertaining to the notion of uncertainty  or indeterminacy are
rather obvious, although they do  not quite  fit to   the current
 quantum information  idea of a "useful" quantum measurement, \cite{brukner}.

Given the  probability density $\rho (x)$  on $R^n$, we define the
differential  entropy  \cite{sobczyk,cover,lasota}),
 as follows:
\begin{equation}
 {\cal{S}}(\rho ) = - \int  \rho (x) \ln \rho (x)\, dx \,
 .
 \label{two}
\end{equation}

One may consider  a subset $\Gamma \subset R^n$  to be a support
of $\rho $ instead of $R$;  this  is guaranteed by the  convention
that the integrand in Eq.~(\ref{two}) vanishes if $\rho $ does.
Note a minor but crucial  notational difference between $\hat{\rho
}$ and $\rho $.

Let us stress that, modulo  minor exceptions, throughout the paper
we carefully avoid dimensional quantities (all relevant
dimensional constants like the Planck $\hbar $  are scaled away),
since otherwise the above  differential entropy definition may be
 dimensionally defective  and have no physical meaning. An extensive
 list od differential entropy values for various probability
 densities can be found in Re. \cite{cover}.

 Since our
paper is supposed to be concerned with physical applications, in
Section II we shall analyze  the issue of how  the differential
entropy definition depends  on the units used. The related
difficulty, often overlooked in the literature, refers to
literally  taking the logarithm of a dimensional argument, see
e.g. \cite{bril,ohya}.

In the quantum mechanical context, we shall
 invoke either position ${\cal{S}}(\rho )$   or momentum ${\cal{S}}(\tilde{\rho })$
  information entropies, with no recourse  to the classical  entropy given in
terms of classical phase-space distributions $f(q,p)$ or (Wehrl
entropy)  their Wigner/Husimi  analogues, \cite{wehrl,halliwell}.
The   notion of  entropic uncertainty relations,
\cite{petz,mycielski,madajczyk,uffink}
 explicitly relies on the differential  entropy input.

 Namely, an  arithmetic  sum  of (presumed to be finite)  momentum and position information entropies
 for   any  normalized  $L^2(R^n)$   wave packet  $\psi (x)$,  is bounded from below:
\begin{equation}
 {\cal{S}}(\rho ) + {\cal{S}}(\tilde{\rho }) \geq  n(1+\ln \pi ) \label{one}
 \end{equation}
 where $n$ stands for
the configuration space (respectively momentum space) dimension,
\cite{mycielski}.
 This feature  is worth emphasizing, since neither
${\cal{S}}(\rho )$ nor  ${\cal{S}}(\tilde{\rho })$   on their own
are   bounded from below or from above. Nonetheless, both take
finite values in physically relevant situations and their sum is
always positive.

\subsection{Temporal behavior-preliminaries}

Since  a normalized  wave function $\psi $  represents a pure
state of a quantum system whose dynamics is governed by the
Schr\"{o}dinger equation, only for stationary states  the
differential  entropy ${\cal{S}}(\rho )$  is  for sure a conserved
quantity. In general, the Schr\"{o}dinger picture  evolution of
$\psi(x,t)$ and so  this of   $|\psi (x,t)|^2 \doteq  \rho (x,t)$
may give rise to  a nontrivial   dynamics  of the  information
entropy associated with   the wave packet  $\psi (x,t)$.

Let us point out that most of the "entropic" research pertains to
time-independent situations, like in case of stationary solutions
of the Schr\"{o}dinger equation. Notable exceptions are
Refs.~\cite{ruiz,majernik,majernik1}.  On general non-quantum
grounds
 an information  (differential entropy) dynamics is addressed in
 Refs.~\cite{sobczyk,mackey}  and \cite{nicolis}-\cite{hatano}, see also
 \cite{qian,qian1,qian2,igarashi,ruelle}.

  The differential entropy, by a  number of
reasons \cite{shannon,sobczyk}, is   said  not to  quantify the
absolute "amount of information carried by   the state of the
system" (Shannon's
  uncertainty), unless carefully interpreted.
  Up to measure preserving coordinate transformations the latter
  objection remains invalid and this feature  gave some impetus to
  numerically assisted  comparative  studies of  the Shannon
  information content of different pure  states of a  given
  quantum system.

Results are  ranging from simple atoms to molecules, nuclei,
  aggregates of particles, many-body Bose and Fermi  systems,
   and Bose-Einstein condensates,  see e.g. Refs.~\cite{gadre} - \cite{balian}.
 In these cases,  Shannon's differential  entropy  appears to be a fully adequate measure
   for the  localization degree (which in turn is interpreted as   both the uncertainty
 measure and the
  information content) of the  involved wave packets.

A difference of two information entropies (evaluated with respect
to the same coordinate system)
 $ {\cal{S}}(\rho ) -  {\cal{S}}(\rho ')$ is known
to  quantify an  absolute change in the  information  content when
passing  from one state of a given  system to another. All
potential problems with dimensional units would  disappear in this
case, \cite{shannon,bril}.
Alternatively, to this end one may
invoke the
 familiar notion of
the relative  Kullback   entropy $-\int_{\Gamma } \rho \, (\ln
{\rho } - \ln {\rho '})\, dx$, \cite{shannon,sobczyk}, provided
$\rho '$ is strictly positive.

 Cogent recommendations  towards
the use of the  Shannon information measure, plainly  against the
Kullback option, can be found in Ref.~\cite{tribus}.  We shall
come to this point later.
 For arguments just to the opposite  see e.g. \cite{smith} and also \cite{tyran,tyran1}.

 In the present paper,  we predominantly invoke the
differential entropy.  In Section IV we shall describe  a number
of limitations upon  the use of the  Kullback entropy.  If both
entropies can be safely employed for the same physical model (like
e.g. for the diffusion type dynamics with asymptotic  invariant
densities), we establish direct  links between the  Shannon and
Kullback entropy dynamics.

  In the context of  the  induced (by time development of probability densities)
    "information dynamics"    ${\cal{S}} \rightarrow  {\cal{S}}(t)$,   \cite{sobczyk,sobczyk1}, it is
 the difference ${\cal{S}}(t) - {\cal{S}}(t')$ between  the (presumed to be finite)   information
 entropy values  for the time-dependent  state of the  same physical system, considered    at  times  $t'<t$,
 which  properly  captures  the  net  uncertainty/information  change  in the respective time
  interval $[t',t]$.  Let us mention that  the very same strategy
  underlies the Kolmogorov-Sinai entropy notion for classical
  dynamical systems, \cite{dorfman,gaspard,lasota}.

 In particular, the  rate in time  of information entropy  $\frac{d{\cal{S}}}{dt}$    is a  well defined
 quantity characterizing    the temporal  changes  (none, gain or loss) in  the  information  content
 of  a given $L^2(R^n)$  normalized wave packet  $\psi(x,t)$ (strictly speaking, of the related
 probability density).
 We indicate  that, at variance with standard thermodynamical   intuitions,  quantum mechanical  information
(differential)  entropy needs not to be a monotonic function of
time. In the course of its evolution it  may  oscillate, increase
 or  decrease  with the flow of time,  instead of  merely  increasing  with time or staying
 constant, as customarily expected.  That, regardless from the intrinsic time reversal property
 of the quantum dynamics.

To conform with the information theory lore, we need to address an
information entropy balance in the course of time, since for  an
isolated quantum  system there is no  analog of  thermal
 reservoir,   capable of  decreasing  (removal) or increasing (putting into)
an  information entropy of the  particular state in which the
system actually is. Since there in no quantifiable  energy
exchange with the environment, the actual purpose and  "fate"  of
the differential entropy need to be investigated.

The entropic method we follow in the present paper extends to  any
formalism operating with general time-dependent  \it spatial \rm
probability densities, \cite{mackey,lasota,nelson}, even if  set
out of the   explicit  thermodynamic context (e.g. the phase space
formulation of  statistical mechanics).

Information entropy and its intrinsic dynamics, like e.g. the
information flow and  information entropy production rate,
quantify properties of  general  reversible and/or irreversible
dynamical systems. Normally, the   microscopic dynamics of such
systems  is expected to follow  well defined trajectories
(deterministic paths of a  dynamical system  or sample paths of a
stochastic process) and those  may  be thought to induce a
corresponding dynamics  for statistical ensembles of trajectories.

It is seldom possible to have a sharp wisdom of the initial data
$x_0 \in X$  for the trajectory dynamics taking place in a phase
space $X$ of the system. This imprecision extends  to the terminal
data ($x_0 \rightarrow x_t$  after time $t>0$) as well.

 Therefore, even if  one knows  exact  dynamical rules governing the behavior
of individual trajectories in time,  it is basically impossible to
tell more about the system then: if its  initial state can be
found in a subset $A \subset X$  with a probability $prob(x_0 \in
A)$, then  after time $t$ one  can identify  the  terminal state
of the system $x_t \in X$ in a subset $B\subset X$ with a
probability  $prob(x_t \in B)$. An evolution of  \it derived \rm
probability densities   eventually  may be obtained  as a solution
of an appropriate  partial differential transport equation,
\cite{mackey,lasota,nicolis,grigolini}

In the present paper we take a more general view  and  we  bypass
a concept of the underlying
 trajectory  dynamics  by  emphasizing the role of  transport equations  and
 their  density solutions.
Under such premises,  we  can safely   address the  \it  dynamics
of  uncertainty/information  \rm
 generated by the Schr\"{o}dinger picture  quantum  evolution of wave packets in closed (\it no \rm
system - reservoir/environment coupling)   quantum mechanical
systems.

{\bf Remark 1:}  Keeping in touch with  quantum mechanical
tradition, let us recall that at least two different "trajectory
pictures "  can be related to the very same mathematical model
based on the Schr\"{o}dinger wave packet dynamics:  deterministic
Bohmian   paths  \cite{durr}  and  random paths of (basically
singular)  diffusion-type processes,  \cite{nelson,carlen,eberle}.
Additionally, under suitable restrictions (free motion,
 harmonic attraction) \it  classical \rm  deterministic   phase-space  paths are
 supported by the associated with $\psi(x,t)$  positive    Wigner distribution function and its spatial
 marginal distribution.
 However,  none of the above  \it   derived \rm  trajectory  "pictures" deserves the status of
 an underlying physical "reality" for quantum phenomena although  each of them  may serve as an adequate
 pictorial description of the  wave-packet dynamics.

{\bf Remark 2:}  In view of   Born's  statistical interpretation
postulate, the Schr\"{o}dinger picture
 dynamics  sets  a well  defined   transport problem for  a probability density
 $\rho (x,t) \doteq |\psi  (x,t)|^2$.  Therefore,   one is   tempted to resolve such dynamics  in terms
 of (Markovian) diffusion-type processes  and their sample paths, see e.g.
 \cite{nelson,carlen,eberle} and  \cite{gar,gar1}.
A  direct  interpretation in terms of random   "trajectories"  of
a Markovian diffusion-type process is here
 in principle possible  under a number of mathematical restrictions, but  is non-unique and not necessarily
 global in time. The nontrivial boundary data,  like the presence of  wave  function  nodes, create
 additional problems  although the nodes  are known to be never reached by the pertinent processes.
 The main source of difficulty lies in guaranteing the existence of a process  per se i.e. of the
well defined transition probability  density function  solving a
suitable parabolic partial
 differential equation (Fokker-Planck or Kramers).
By adopting milder conditions upon the  drift fields (instead of
too restrictive
 growth restrictions, one may simply  admit  smooth functions) it is possible to  construct well defined,
  albeit non-unique,  diffusion-type processes. They  are   consistent with the  time development of a
  given probability density, see Chap. 3 of Ref.~\cite{qian1} and \cite{eberle}.

\subsection{Outline of the paper}

The paper is structured as follows.  We begin by  recalling  the
standard lore of the   Shannon information theory to attribute an
unambiguous  meaning to two principal notions, this of  \it
information \rm and that of \it uncertainty. \rm To this end
various notions of  \it state \rm   of a  model  system are
invoked and suitable information measures are discussed.

Next we  turn to  the coarse-graining issue  and  set a
connection between the  Shannon  entropy of a discrete probability
measure and the differential entropy of
 a  related  (through a suitable limiting procedure)
  continuous probability density.  On the way, we analyze the dimensional
  units impact on the entropy definition.
We discuss various   entropic inequalities  for both differential
and coarse-grained  entropies of    quantum mechanical densities.

 In Section III,  the    localization  degree   of probability densities
  is analyzed by means of  so-called entropy powers and  of the
   Fisher information measure. We infer    two chain inequalities,  Eqs.~(\ref{in1})
   and (\ref{in2}),   which imply that typically
   the differential entropy is  a well behaved quantity,  bounded both
   from below and above.  The formalism is  general  enough to include  quantum
   mechanical densities as merely the  special case.

In Section IV  we set a conceptual framework for time-dependent
problems.
 Since  classical dynamical,  stochastic and quantum systems (in their pure states) in general
 give rise
to time-dependent probability densities and information entropies,
we resolve  the exemplary density dynamics in terms of
Smoluchowski   diffusion processes, albeit   with no explicit
random path (e.g. random variable) input.

The entropy   and Fisher information evolution equations are
established. Close links of the differential and  conditional
Kullback entropies are established for Smoluchowski diffusion
processes, when asymptotic invariant densities enter the scene.
 We discuss  a compliance  of the induced
continual power release in the course of the diffusion process
with the mean energy conservation law, Eqs~(\ref{interplay}) and
(\ref{Fishdynamics}).

In section V we analyze   differential entropy dynamics  and time
evolution of the Fisher localization measure  in quantum theory
and next  exemplify the general
 formalism  for  simple analytically tractable cases.
 The emergent continual  power transfer effect has been analyzed
 in connection  with the finite   energy constraint for the mean energy of
 quantum motion, Eqs.~(\ref{interplay1}) and (\ref{Fishdynamics1}).

Although uncertainty dynamics scenarios of  sections IV and V
 are fundamentally  different, nonetheless the respective  methodologies appear to  have an
 overlap, when restricted  to steady  states which support invariant densities
 for  (reversible) stationary  diffusion-type processes.

\section{Differential entropy: uncertainty versus information}

\subsection{Prerequisites}

The original definition of Shannon entropy  conveys a dual meaning
of both   uncertainty and information measure.  It is useful to
interpret those features in a complementary  (albeit colloquial)
way: the less is the uncertainty of  the system or its state, the
larger (and more valuable)  is the information
 we acquire as a result of the  measurement (observation) upon the system, and in reverse.

We know that  a result of an observation of any random phenomenon
cannot be predicted a priori (i.e. before an observation), hence
it is natural to quantify an  uncertainty of this phenomenon. Let
us  consider $\mu =(\mu _1,...,\mu _N)$ as a probability measure
on  $N$  distinct (discrete)    events $A_j, 1\leq j\leq N$
pertaining to  a  model  system. Assume that  $\sum_{j=1}^{N} \mu
_j = 1$ and  $\mu _j =prob(A_j)$  stands for  a probability for an
event $A_j$ to occur in the game of chance with $N$ possible
outcomes.

 Let us call $ - \log \, \mu _j$ an
 \it uncertainty  function \rm  of the event $A_j$.
Interestingly, we can  coin  here  the name of  the ("missing")
\it information function, \rm  if we wish to interpret what  can
be learned  via  direct observation    of the event $A_j$: the
less probable is that event, the more valuable (larger) is the
information we would retrieve through its registration .

 Then,  the expression
\begin{equation}
{\cal{S}}(\mu ) = - \sum_{j=1}^N \mu _j \log \, \mu _j  \,
\label{info}
\end{equation}
stands for  the  measure of the   \it  mean    uncertainty  \rm
  of the possible outcome of the game, \cite{sobczyk}, and
at the same time quantifies the \it  mean information \rm
 which is  accessible  from an experiment  (i.e.
actually playing the game). The base of the logarithm
 for a while is taken equal $2$, but we recall that $\log \, b \cdot \ln \, 2 = \ln \, b$ and
$\ln 2\simeq 0.69555$ with the  base  $e\simeq 2.71828$  .

Thus, if we identify   event values   $A_1, ..., A_N$  with labels
for   particular
 discrete  "states" of the system, we may interpret  Eq.~(\ref{info}) as a measure of uncertainty
of the  "state"  of the system, \it before \rm   this particular
"state" it is chosen out of the set of  all admissible ones. This
well conforms with the standard  meaning attributed to the Shannon
entropy: it is  a measure of the degree of ignorance concerning
which possibility (event $A_j$) may hold true in the set $\{ A_1,
A_2,...,A_N\}$ with  a given a priori probability distribution $\{
\mu _1, ...,\mu _N\}$.

Notice that:
\begin{equation}
0 \leq  {\cal{S}}(\mu ) \leq  \log \, N  \, \label{uncertain}
\end{equation}
ranges from certainty (one entry whose  probability equals $1$ and
thus no information  is missing)
  to  maximum uncertainty when a uniform distribution $\mu _j = 1/N$ for all $1\leq  j \leq N$  occurs.
 In the latter situation, all events (or measurement outcomes) are equiprobable and $\log \,N$ sets
maximum for a measure of the "missing  information".

By looking at all intermediate levels of randomness allowed by the
inequalities Eq.~(\ref{info})  we realize that
 the lower is  the Shannon entropy the  less information about "states" of the system   we are missing, i.e. we have
more information  about the system.

 If the Shannon entropy  increases,  we actually loose an information available  about the system.
Consequently, the difference between two uncertainty measures can
be interpreted as an information gain or loss.

\subsection{Events, states, microstates and macrostates}

The   Boltzmann formula
\begin{equation}
{\cal{S}}=k_B \ln W \doteq  - k_B \ln P  \label{boltz}
 \end{equation}
 sets a link of  entropy  of the  (thermodynamical)  system with the probability $P=1/W$
that an  appropriate "statistical  microstate" can  occur. Here,
$W$ stands for a number of all possible (equiprobable) microstates
that imply  the  prescribed macroscopic (e.g. thermodynamical)
behavior corresponding to a \it fixed \rm value of ${\cal{S}}$.

It is instructive to recall that  if $P$ is  a probability of  an
event i.e. of a particular microstate, then $- \ln P$  (actually,
with  $\log _2$ instead of $\ln $)  may be interpreted
 \cite{hartley}  as "a measure of information produced when one message
  is chosen from the set, all choices being equally likely" ("message" to be identified
  with a  "microstate").  Another interpretation of
  $- \ln P$ is  that of a degree of uncertainty in the trial experiment, \cite{yaglom}.

 As a pedestrian illustration let us invoke a classic example of a
 molecular gas  in a box which is divided into two halves denoted "$1$" and "$2$". We allow the
 molecules   to be in one of two  elementary  \it  states: \rm
  $A_1$ if a molecule can be found in  "$1$"  half-box and
$A_2$ if it placed in another  half "$2$".

Let us consider a  particular $n$-th  \it macrostate \rm
 of a molecular gas  comprising a  total of $G$ molecules  in a box,  with $n$ molecules
 in the state $A_1$ and $G-n$ molecules in the state $A_2$.

The total number of ways in which  $G$  molecules  can be
distributed between two halves of the box in this prescribed
macrostate,   i.e.  the number $W = W(n)$   of distinct
equiprobable  \it microstates, \rm clearly is $W(n) = G!/[n!
(G-n)!]$. Here,  $P(n) =1/W(n)$ is a probability with which any of
microstates may occur in  a system  bound to "live"   in a given
macrostate. The maximum of $W(n)$  and thus of $k_B\ln W(n) $
corresponds to $N_1=N_2=n$.

To get a better insight into the  information-uncertainty
intertwine, let us  consider an ensemble of  finite systems which
are  allowed to appear  in  any of  $N>0$ distinct elementary
states. The meaning of "state" is left unspecified, although an
"alphabet" letter may be  invoked for convenience.

Let us  pick up   randomly  a large sample composed of  $G \gg 1$
single systems, each  one in a certain  (randomly assigned)
state. We  record  frequencies $n_1/G \doteq p_1,...,n_N/G \doteq
p_N$ with which the  elementary  states  of the type $1,...,N $ do
actually occur.
 This sample is a substitute for a "message" or a "statistical microstate" in the previous
 discussion.

Next, we   identify the number of \it  all \rm  possible  samples
of that    fixed  size $G$  which  would   show up    the very
same statistics $p_1,...,p_N$  of elementary states. We interpret
those samples to display  the same   "macroscopic behavior".

  It was the major discovery due to
  Boltzmann, see e.g.  \cite{shannon},  that  the   number  $W$  of  relevant  "microscopic   states"
   can be approximately  read out from  each single   sample   and  is  directly  related
   to the the  introduced a priori  probability measure $\mu _1,...,\mu _N$, with an identification
    $p_i \doteq \mu _i$ for all   $1\leq i\leq N$,   by the  Shannon
   formula:
 \begin{equation}
  \ln W \simeq   - G  \sum_{i=1}^N  p_i \ln p_i  \doteq  - G \cdot {\cal{S}}(\mu )  \label{shan}
 \end{equation}

On the basis of this formula, we  can consistently introduce
${\cal{S}}(\mu )$ as   the \it   mean information \rm   per each
 ($i$-th)  elementary state of
the $N$-state    system, as encoded in a given  sample  whose
size
  $G\gg 1$ is sufficiently large,  \cite{bril}.

 To exemplify previous considerations, let us  consider $N=2$.
  It is instructive to  compare the uncertainty level (alternatively - information content)
of ${\cal{S}}(\mu )$ for the two-state system, if we  take $2$ as
the logarithm base instead of $e$. Then, Eq.~(\ref{shan}) would
refer  to the  binary encoding of the  message (string) with $G$
entries.

We find that  $\mu _1 =0.1$ and  $\mu _2=0.9$  yield
${\cal{S}}(\mu ) = 0.469$. Analogously
  $0.2$ and  $0.8$ imply $0.7219$, while $0.3$ and  $0.7$ give $0.8813$.
  Next,  $0.4$ and  $0.6$ imply $0.971$, and we reach  an obvious
maximum ${\cal{S}} =1$ for $\mu _1=\mu _2= 0.5$.  An instructive
example of the "dog-flea" model workings with $G=50$ fleas jumping
back and forth
 between their "states of residence" on a dog "$1$" or dog "$2$",  can be found in
 Ref.~\cite{dog}.  Albeit, in a number of specific cases, an evolution of the Gibbs entropy may
 show up some surprises if the "entropy growth dogma" is
 uncritically accepted, see e.g. examples in \cite{dog,sobczyk1}
 and the discussion of Refs.~\cite{huang,cercignani}.

By pursuing the Shannon's communication theory track,
\cite{shannon}, we can as well identify states of the model system
with  "messages" (strings)  of  an  arbitrary  length $G >0$
which  are entirely
 composed by  means of the prescribed  $N$  "alphabet" entries (e.g. events or alphabet letters $A_j$
with the previous probability measure $\mu $). Then,
Eq.~(\ref{shan}) may  be interpreted as  a measure of \it
information \rm   per alphabet letter,  obtained \it after \rm  a
particular message (string $\equiv $ state of the model system)
has been received or measured, c.f. our discussion preceding
Eq.~(\ref{shan}).  In this case, the
 Shannon entropy interpolates between a maximal information  (one certain event)
and  a minimal information (uniform distribution), cf.
Eq.~(\ref{uncertain}).\\

 {\bf Remark 3:} Any  string containing $G=10.000$ symbols which are  randomly sampled  from
among  equiprobable  $N=27$ alphabet letters, \cite{bril}, stands
for a concrete microstate.  In view of
 $\mu =1/27$, a  corresponding   macrostate   is described  via Eqs. ~(\ref{info}) and (\ref{shan})
 in terms of the  number  ${\cal{S}}(\mu ) =
  - \log_2 (1/27) \simeq 4.76$.
  Accordingly,    $ \log _2 W = G\cdot {\cal{S}}(\mu ) \simeq 47.600$,
   where $W$ is  the number of admissible  microstates.

\subsection{Shannon entropy   and differential entropy}

\subsubsection{Bernoulli scheme  and  normal distribution}

Let us consider again a two-state  system where $A_1$ appears with
a probability $\mu _1=p$ while $A_2$ with a probability $\mu _2= 1
- p$. A probability with which $A_1$ would  have appeared exactly
$n$ times, in the  series of $G$ repetitions of the two-state
experiment,  is given by the Bernoulli formula:
\begin{equation}
P_n = {\frac{G!}{[n!(G-n)!]}} p^n(1-p)^{G-n} \label{ber}
\end{equation}
where, in view of the  Newton   formula for the binomial
$(p+q)^G$, after  setting  $q=1-p$ we  arrive  at $\sum_{n=0}^{G}
P_n = 1$.

Since the number $n$  of successes in the  Bernoulli scheme is
restricted only by  inequalities $0\leq n\leq G$,  what  we have
actually  defined  is a probability measure  $\mu = \{
P_0,P_1,...,P_G \} $ for $G$  distinct random   events denoted
$B_0,B_1,...,B_G$. Accordingly,     we    can introduce   a random
variable $B$ and say that it has the Bernoulli distribution, if
$B$  takes values $n=0,1,...,G$ with the Bernoulli  probabilities
$P_n$ of Eq.~(\ref{ber}) for all $n$. A random  event $B_n$ is
interpreted as "taking the value $n$ in the Bernoulli scheme".

Let us denote $P(B=k)  \doteq P(B_k) =  P_k$.  We know that $P(B<
n) = \sum_{k<n} P_k$.  The mean value $E(B)$ of $B$ reads
$E(B)=\sum_{k=0}^G  n \cdot P_k = G p$. The variance
$E([B-E(B)]^2)$  of $B$ equals  $Gp(1-p)$.

The local de Moivre-Laplace theorem tells  us  that  for large
values of $G$ the binomial distribution can be
 approximated  by the normal  (Gauss) one:
\begin{equation}
P_n \equiv  {\frac{1}{[2\pi Gp(1-p)]^{1/2}}}\, \exp \left( -
{\frac{(n-Gp)^2}{2Gp(1-p)}}
 \right)  \, .  \label{gauss}
\end{equation}

At this point we shall take an inspiration from
Ref.~\cite{chandra} and relate the Bernoulii "success"
probabilities  with probabilities of locating a particle in an
arbitrary  interval on a line $R$. Namely, let  us first  consider
an interval of length L: $[0,L]\subset R$. Let $G \gg 1$, we
define  an interval grating unit to be $ r = L/G$ and next
redefine  Eq.~(\ref{gauss}) to  arrive at a  probability per bin
of length $r \ll 1 $:
\begin{equation}
{\frac{1}{r}} P_n = \rho (x_n)  = {\frac{1}{[2\pi \sigma
^2]^{1/2}}} \exp\left( -{\frac{(x_n - x_0)^2} {2\sigma ^2}}
\right) \label{gauss1}
\end{equation}
with: $x_n = nr$, $x_0=Gpr$ and $\sigma ^2 = Gr^2 p(1-p)$.
Obviously, $\rho (x_n)$ is  not a probability on its own, while
$r \cdot  \rho (x_n) = P_ n$ \it  is \rm  a probability to  find a
particle in the $n$-th interval of length $r$ out of the admitted
number $G = L/r$ of bins.

For convenience let us specify $p=1/2$  which implies  $x_0= r G/2
$ and $\sigma = r \sqrt{G}/2 $. We recall that almost all of  the
probability  "mass" of the Gauss distribution  is  contained in
the interval $-3\sigma < x_0 < +3\sigma $ about the mean value
$x_0$. Indeed, we have $prob(|x- x_0|)<  2\sigma ) = 0.954$  while
$prob(|x- x_0|)<  3\sigma ) = 0.998$.

The  lower bound $100<n \leq G $  justifies the usage of
simplified versions of the standard
 Stirling formula    $ n^n\exp(-n) \sqrt{2\pi n} < n! <   n^n\exp[-n  + (1/12n)]  \sqrt{2\pi n}$,
 in view of the above   probability "mass" estimates. Therefore,  we can
 safely replace the Bernoulli probability measure by  its (still discrete) Gaussian approximation
  Eq.~(\ref{gauss}) and next pass to  Eq.~(\ref{gauss1}) and its obvious continuous
  generalization.

{\bf  Remark 4:} By taking a concrete  input of  $L=1$ and
$G=10^4$ we get the
 grating (spacing, resolution) unit  $r=10^{-4}$. Then, $x_0= 1/2$ while $ \sigma = (1/2) \cdot 10^{-2}$.
 It is thus  a  localization  interval $[1/2 - 3\sigma ,1/2+ 3\sigma ]$ of
 length $6 \sigma = 3\cdot
  10^{-2}$ to be compared with  $L=1$.  By setting $G=10^6$ we would get
   $6\sigma =  3\cdot 10 ^{-3}$.

    For future reference, let us stress  that generally we  expect
     $r \ll \sigma$ which implies a sharp distinction between the grating
     (resolution) unit $r$  and  the localization properties of the Gauss
     function expressed through its half-width   $\sigma $.

\subsubsection{Coarse-graining}

For a given  probability density function on $R$ we can adopt the
coarse-graining procedure, \cite{mackey}, giving   account of an
imprecision with which a  spatial position $x$ can be measured or
estimated. Thus, if compared with the previous Bernoulli
$\rightarrow $ Gauss  argument,  we  shall proceed
 in reverse from density functions to approximating them  by   piece-wise constant,
  histogram-type discontinuous functions.

We need to partition the configuration space $R$ into a  family of
disjoint  subsets (intervals) $\{ B_k \}$ such that $\cup _k B_k
\subseteq R$ and $B_i \cap  B_j =$\O   \,  for $i\neq j$.
 We denote
 $\mu(B_k)\doteq \mu _k$ the
length of the $k$-th  interval, where $\mu $ stands for the
Lebesgue measure on $R$.

A probability that  a Gaussian random variable  with the density
$\rho $ takes its value $x$ in  an interval $B_k$   equals
$prob(B_k) \doteq p_k =  \int_{B_k} \rho (x) dx$. An average of
the density $\rho $ over $B_k$ we denote  $<\rho >_k =   p_k / \mu
_k$ where $\mu _k= \int_{B_k} dx$.

The probability density $\rho $ coarse grained with respect to the
partition $\{ B_k \} $ reads:
\begin{equation}
\rho _B(x)  \doteq  \sum_k <\rho >_k 1_k(x)    \label{coarse}
\end{equation}
where $1_k(x)$ is an indicator (characteristic) function of the
set $B_k$, which is
 equal $1$ for $x\in B_k$ and
 vanishes  otherwise.
 Since  $\int 1_k(x)  dx = \mu _k$ it is clear that
\begin{equation}
  \int \rho _B(x) dx =  \sum_k  <\rho >_k \mu _k = \sum_k p_k =  1  \label{coarse1}
 \end{equation}
where an interchange of  the summation with integration is
presumed to be allowed.

By invoking arguments  of the previous subsection, we choose  a
grating  unit  $\mu _k = r\ll 1$ for all $k$ and notice that
 $<\rho >_k = p_k/r$ where   $p_k \simeq  \rho (x_k) \cdot r$
 for certain   $x_k \in B_k$.

In view of the  twice  "triple half-width"  spatial localization
property of the Gauss function,   we can safely assert  that  an
interval $L\sim 6\sigma $ about $x_0$ may be used  in the coarse
graining procedure, instead of the full  configuration space $R$.
Effectively, we arrive at a \it  finite \rm partition on  $L$
with the resolution
 $L/G = r$ and then  we
can safely   invoke the definition of $p_k \doteq P_k = r\cdot
\rho (x_k) $,
 in conformity with  Eq.~(\ref{gauss1}).

For a coarse grained  probability  density we  introduce a coarse
grained Shannon entropy whose  relationship to the original
differential entropy is of major interest. We have:
\begin{eqnarray}
{\cal{S}}(\rho _B) =  - \sum_k  p_k \ln p_k  \simeq & &   \\
- \sum_k [r \rho (x_k)] \ln  r  -  \sum_k [r \rho (x_k)] \ln
[\rho  (x_k)] & \nonumber
    \label{limit}
\end{eqnarray}
with a standard interpretation of the  mean  information  per bin
of length $r$. Here,  if  a partition (grating) unit $r$ is
small, one  arrives  at  an approximate formula  (we admit $|\ln
r| \gg 1$):
\begin{equation}
 {\cal{S}}(\rho _B) \simeq    - \ln r - \int \rho (x) \ln [\rho (x)] dx  =
 -  \ln r  + {\cal{S}}(\rho )
\,  \label{grating}
\end{equation}
with the obvious proviso that ${\cal{S}}(\rho _B) \geq 0$ and
hence, in view of
  ${\cal{S}}(\rho ) \geq  \ln r$,    we need to have   maintained
    a proper balance between
$\sigma $ and the chosen grating level $r$.

{\bf Remark 5:}  It is instructive to validate the above
approximation for the choice of  $r=10^{-6}$, hence $\ln r   = - 6
\ln 10 \sim -13.86$ . We have: $ {\cal{S}}(\rho ) =  (1/2) \ln
(2\pi e \sigma ^2) \sim 0.92 + \ln \sigma $. By setting $\sigma =
(1/2)10^{-3}$ we realize that ${\cal{S}}(\rho ) \sim 0.92 -\ln 2 -
3 \ln 10 \sim  - 6.7$  hence is  almost twice larger than  the
allowed lower bound
  $\ln r$.
The approximate value of the coarse grained entropy is here
${\cal{S}}(\rho _B) \sim  7.16$ and stands for the mean
information per partition bin in the "string" composed  of $G$
bins.

In view of Eq.~(\ref{grating}),  as long as we keep in memory the
strictly positive
 grating unit $r$, there is a well defined "regularization"   procedure   (add $-\ln r$
 to  ${\cal{S}}(\rho )$)
 which relates   the coarse grained entropy  with  a given differential entropy.
In a number of cases it is computationally simpler to evaluate the
differential entropy, and then  to  extract  - otherwise
computationally  intractable - coarse grained entropy.

Notice that one cannot allow  a naive zero grating limit in
Eq.~(\ref{grating}), although  $r$  may  be arbitrarily small.
Indeed,  for  e.g. the  Gaussian densities, differential entropy
takes  finite values and
 this would suggest that the coarse grained entropy might be arbitrarily large.
This obstacle does not arise  if one proceeds with some care. One
should remember that we  infer the coarse grained entropy from the
differential entropy exactly at the price of introducing the
resolution  unit $r$. The smaller is $r$, the better is an
approximation of the differential entropy by the second term on
the right-hand-side of Eq.~(\ref{limit}), but $-\ln r$ needs to
remain as a finite entry in Eq.~(\ref{grating}).

We have  inequalities  $0 \leq {\cal{S}}(\rho _B) \leq G$  where
$L= G \cdot r$. They extend to all  approximately equal  entries
in  Eq.~(\ref{limit}). Since $-\ln r = - \ln L + \ln G $, we
arrive at  new  inequalities:
\begin{equation}
\ln r   \leq   - \sum_k [r \rho (x_k)] \ln  [\rho  (x_k)]  \leq
\ln L
\end{equation}
where $\sum_k [r \rho (x_k)] \ln  [\rho  (x_k)] \Rightarrow
   - \int \rho \ln \rho \, dx $ with  $r\rightarrow 0$ and possibly $L\rightarrow \infty $.
A conclusion is that the differential entropy is  unbounded both
form below and from the above. In particular,   ${\cal{S}}(\rho )$
may take arbitrarily low  negative values,   in  plain  contrast
to  its coarse grained version  ${\cal{S}}(\rho _B)$ which
  is always nonnegative.

We can be more detailed  in connection with  approximations
employed in   Eqs.~($16$)
 and (\ref{grating}).  Actually, the right-hand-side of Eq.~(\ref{grating}) sets
 a lower bound for the coarse-grained entropy ${\cal{S}}(\rho _B)$.

 Let us  recall that the value of a convex function $x\ln x$  at  the mean value of
 its argument $\langle x\rangle $, does not exceed the mean value
  $\langle x\ln x\rangle $  of the function itself. Then, in our notation which follows
  Eq. (12),  we  can directly employ an averaging over $B_k$:
\begin{equation}
{\frac{1}{r}} \int_{B_k} \rho  \ln \rho  dx  \geq \left(
{\frac{1}{r}} \int_{B_k} \rho  dx \right) \left[ \ln \left(
{\frac{1}{r}}  \int_{B_k} \rho dx \right) \right] \, .
\end{equation}

Taking the minus sign,   executing summations with respect to $k$
(convergence of the series being  presumed) and using Eqs. (15)
and (16)  we get:
\begin{equation}
{\cal{S}}(\rho )  -  \ln \, r  \leq  {\cal{S}}(\rho _B) \,
\label{ineq}
\end{equation}
as a complement  to   Eq. (17), see e.g. also
\cite{ruiz,madajczyk}.

Equations (16) and (\ref{ineq}) allow, with suitable reservations,
to extend the standard information/uncertainty measure meaning
from coarse-grained entropies to  differential entropies per se.
Namely, the difference of two coarse grained  entropies,
corresponding to the same  partition but to different (coarse
grained) densities, may be adequately  approximated by the
difference of the corresponding differential entropies:
\begin{equation}
 {\cal{S}}(\rho _B) - {\cal{S}}(\rho '_B) \simeq {\cal{S}}(\rho ) - {\cal{S}}(\rho ') \, , \label{appr}
 \end{equation}
 provided they take finite values, \cite{sobczyk,ruiz}:

 \subsubsection{Coarse-graining exemplified: exponential density}

An exponential density on a positive half-line $R^+$  is known to
maximize a differential entropy among all  $R^+$ density functions
with the first moment fixed at  $1/\lambda >0 $. The density has
the form: $\rho (x) = \lambda  \exp (- \lambda x)$ for $x\geq 0$
and vanishes for $x<0$. Its variance is $1/ \lambda ^2$. The
differential entropy of the exponential density reads
${\cal{S}}(\rho ) = 1 - \ln \lambda $.

In he notation of the  previous subsection, let us coarse-grain
this density  at a particular value of $\lambda =1$ and  then
evaluate  the corresponding entropy as follows.
 We choose $r\ll 1$, $p_k \simeq \rho (x_k)
\cdot r$ with $x_k = k r $ where $k$ is a natural number. One can
directly verify that for small $r$, we can  write $r\simeq
1-exp(-r)$ and thence consider $p_k \simeq  [1-exp(-r)] \exp(-k
r)$, such that $\sum_{k=0}^{\infty } p_k =1$, with
 the well known quantum mechanical connotation.

 Namely, let us  set $r = h\nu /k_BT$; notice that  for the first
 time in the present paper, we  explicitly invoke dimensional units,
 in terms of which the dimensionless constant  $r$ is defined.
 We  readily  arrive at the probability of the $k \nu $-th
oscillator mode  in thermal bath  at the temperature $T$.

Our assumption of $r\ll 1$ corresponds to the low frequency
oscillator problem with $\nu \ll 1$, see e.g. \cite{bril}.
Clearly, we have for the mean number of modes
\begin{equation}
\langle n \rangle  = \sum kp_k = {\frac{1}{\exp r - 1}} \simeq
{\frac{1}r}
\end{equation}
which  implies the  familiar Planck  formula
\begin{equation}
\langle E \rangle = {\frac{h\nu }{ \exp(h\nu /k_BT) - 1}}  \simeq
h\nu /r =k_BT \, .
\end{equation}

 For  the variance we get
\begin{equation}
\langle (n-\langle n \rangle )^2 \rangle = \langle  n\rangle ^2 -
\langle n \rangle \simeq {\frac{1}{r^2}} \, .
\end{equation}

The standard Shannon entropy of the discrete probability
distribution $p_k, k \in N$ reads
\begin{equation}
{\cal{S}}(\rho _B) = - \sum p_k \ln p_k  = - \ln[1-\exp(-r)] +
\langle n \rangle r
\end{equation}
so that in  view of  ${\cal{S}}(\rho _B)  \simeq - \ln r  + 1$
and ${\cal{S}}(\rho )=1 $ for $\rho (x)= \exp (-x)$, we clearly
have ${\cal{S}}(\rho _B)  \simeq - \ln r   + {\cal{S}}(\rho )$, as
expected.

Let us point out our quite redundant sophistication.   In fact,
  we could have skipped all the above   reasoning and take Eq. ~(\ref{grating}) as the starting
  point to evaluate ${\cal{S}}(\rho _B)$ for the coarse grained  exponential
   density,    at the assumed resolution $r= h\nu /k_BT \ll 1$,
 with the obvious   result
\begin{equation}
  {\cal{S}}(\rho _B)\simeq 1- \ln (h\nu /k_BT) \gg 1\, .
\end{equation}

{\bf Remark 6:}  An  analogous procedure can be readily adopted
to analyze
 phenomenological   histograms  of energy spectra for  complex  systems, with an obvious
  extension of  its range of validity to spectral properties
   of the  classically chaotic case. In particular,  semiclassical  spectral series
  of various quantum systems  admit  an approximation of spacing histograms by
  continuous distributions on the positive half-line $R^+$.
Although a full fledged  analysis is quite complicated, one may
invoke quite useful, albeit
 approximate formulas  for adjacent level spacing distributions.
 The previously mentioned exponential density   corresponds to
 the so-called Poisson spectral series.  In  the  family of radial densities of the
form $\rho _N(x) =  {2\over {\Gamma (N/2)}} x^{N-1} exp(-x^2)$
where $N>1$ and $\Gamma $
 is the Euler gamma function, \cite{acta}, the particular   cases $N=2,3,5$
 correspond to the  generic level spacing distributions, based on the exploitation of the
 Wigner surmise. The respective histograms plus their continuous density interpolations
   are often  reproduced in "quantum chaos" papers,  see  for example \cite{well}.

\subsubsection{Spatial coarse graining in quantum mechanics}

The coarse grained entropy  attributes the  "mean information per
bin  of length  $r$" to systems described by continuous
probability densities and  their  differential entropies.
Effectively one has a tool which allows to accompany the coarse
grained density    histogram  (of $p_k$ in the $k$-th  bin on $R$)
by the  related  histogram of  uncertainties $-\ln p_k$, c.f.
Section II.A where an uncertainty function has been introduced.

The archetypal example of position measurement in quantum
mechanics presumes  that position is measured in bins
corresponding to the resolution of the measurement apparatus. This
means that  the continuous spectrum of the position observable is
partitioned into a countable set of intervals (bins) whose
maximum length we regard  as a "resolution unit". For an interval
$B_k\subset R$ we may denote $p_k$ the probability of finding the
outcome of a position measurement to have a value in $B_k$.
  We  are free set the bin size arbitrarily, especially  if computer assisted procedures
  are employed, \cite{rabitz}.

Following \cite{deutsch} one  may take the view  that the most
natural measure of the uncertainty in the result of a  position
measurement or preparation  of  specific localization features  of
a state vector (amenable to an analysis via spectral properties of
the position operator) should be the
 information entropy coming from Eqs.~({\ref{coarse}) and (\ref{coarse1}):
${\cal{S}}(\rho _B) \doteq - \sum_k p_k \ln p_k $ with  a direct
quantum input
 $p_k \doteq \int_{B_k} |\psi (x)|^2 dx$, where  $\psi \in L^2(R)$ is  normalized.
This viewpoint is validated by  current experimental techniques in
the domain of matter wave interferometry, \cite{zeilinger,rauch},
and  the associated  numerical experimentation where various
histograms are generated, \cite{rabitz}.

The  formula Eq.~(\ref{grating}) gives meaning to the intertwine
of  the differential  and coarse grained
 entropies in the quantum mechanical context. When an  analytic  form of the entropy  is in the reach,
 the coarse graining is straightforward. One should realize that   most of the results known to date
 have been obtained numerically, hence with an  implicit coarse-graining, although  they were
  interpreted in terms of the differential entropy, see e.g. \cite{halliwell}-\cite{massen}.

In connection with an entropic  inequality  Eq.~(\ref{one}) let us
point out  \cite{petz}
 that it is a generic property of normalized   $L^2(R^n)$ wave functions  that,
  by means of the Fourier transformation,    they give rise to
  two interrelated  densities (presently we refer to $L^2(R)$):
  $\rho =|\psi |^2$  and $\tilde{\rho } = |{\cal{F}}(\psi )|^2$
where
\begin{equation}
({\cal{F}}\psi )(k) = {\frac{1}{\sqrt{2\pi }}} \int \psi (x) \exp
(-ikx)\,  dx
\end{equation}
 is the Fourier transform of $\psi (x)$. The inequality    (\ref{one}) for the
 corresponding  (finite) differential  entropies  follows,  here  with  $n=1$.

 By choosing  resolutions $r\ll 1$ and $\tilde{r}\ll 1$ we can introduce the respective coarse
 grained  entropies,
 each fulfilling an inequality Eq.~(\ref{ineq}). Combining these inequalities with Eq.~(\ref{one}), we get
 the prototype entropic inequalities for coarse  grained entropies:
 \begin{equation}
 {\cal{S}}(\rho _B) + {\cal{S}}(\tilde{\rho }_B) \geq 1 + \ln \pi  - \ln (r\cdot \tilde{r})
 \end{equation}
with the  corresponding  resolutions $r$ and $\tilde{r}$.

By referring to Eq.~(\ref{grating}) we realize that  the knowledge
of
 ${\cal{S}}(\rho _B)$,  completely determines ${\cal{S}}(\tilde{\rho }_B)$ at the
 presumed resolution levels:
\begin{equation}
{\cal{S}}(\tilde{\rho }_B) \simeq  1 + \ln \pi  - \ln (r\cdot
\tilde{r})  -{\cal{S}}(\rho _B) \geq 0
\end{equation}
and in reverse.
 This in turn  implies that in all computer
  generated  position-momentum differential entropy inequalities,  where the coarse graining
  is implicit,   the knowledge of  position entropy
  and of the resolution levels provide  sufficient data to deduce   the combined
   position-momentum outcomes,
  see also  \cite{stotland}-\cite{massen}.

In standard units (with $\hbar  $ reintroduced), the previous
discussion pertains to quantum mechanical  position - momentum
entropic uncertainty relations. In the notation of Refs.
\cite{partovi1} and \cite{madajczyk} we have:
\begin{equation}
S^x + S^p \geq 1 - \ln 2 - \ln \left( {\frac{\delta x \cdot \delta
p}{h}} \right)             \label{unit}
\end{equation}
 for measurement entropies with  position and momentum  "abstract measuring device"  resolutions $\delta x$
and $\delta p$ respectively, such that    $\delta x \cdot \delta p
\ll h$.

Note,  that to reintroduce $\hbar $ we must explain how the
differential entropy definition is modified if we pass to
dimensional units, see e.g next subsection.  Let us also  point
out that one should not confuse the above resolution units $r,
\tilde{r}$ and $\delta x,\delta p$  with standard mean square
deviation values  $\Delta X$ and $\Delta P$  which are present in
the canonical indeterminacy   relations: $\Delta X \cdot \Delta P
\geq \hbar/2 $.

 Let us set $\hbar \equiv 1$ (natural units system).  If, following conventions we
 define the squared  standard
deviation (i.e. variance)  value for an observable $A$ in a pure
state $\psi $ as $(\Delta A)^2 = (\psi , [A - \langle A\rangle ]^2
\psi )$ with $\langle A \rangle = (\psi , A\psi)$, then for the
position $X$ and momentum $P$ operators we have the following
version of the entropic uncertainty relation  (here expressed
through so-called entropy powers, see e.g.   \cite{petz}):
\begin{equation}
\Delta X \cdot \Delta P \geq  {\frac{1}{2\pi e}} \,
 \exp[{\cal{S}}(\rho )  + {\cal{S}}(\tilde{\rho })] \geq {\frac{1}{2}} \label{heisenberg}
\end{equation}
which is  an alternative for Eq.~(\ref{one}); $n=1$  being
implicit.

\subsection{Impact of dimensional units}

Let us come back to an issue of reintroducing physical units in
Eq.~(\ref{unit}). In fact, if $x$ and $p$ stand for
one-dimensional phase space labels and $f(x,p)$ is a normalized
phase-space density, $\int  f(x,p) dx dp=1 $, then the related \it
dimensionless \rm differential entropy is defined as follows,
\cite{ohya,dunkel}:
\begin{equation}
{\cal{S}}_h = - \int (h f) \ln (h  f) {\frac{dx dp}h}  = - \int f
\ln (h f) dx dp
\end{equation}
where $h=2\pi \hbar $ is the Planck constant. If $\rho  (x)= |\psi
|^2(x)$, where $\psi \in L^2(R)$, then $\tilde{\rho }_h(p) =
|{\cal{F}}_h(\psi )|^2(p)$  is defined in terms of the dimensional
Fourier transform:
\begin{equation}
({\cal{F}}_h\psi )(p) = {\frac{1}{\sqrt{2\pi \hbar }}} \int \psi
(x) \exp (-ipx/\hbar)\,  dx \, .
\end{equation}
Let us consider the joint density
\begin{equation}
f(x,p) \doteq \rho (x) \tilde{\rho }_h(p)
\end{equation}
and evaluate the differential entropy   ${\cal{S}}_h$ for this
density.  Remembering that  $\int \rho  (x) dx =1=  \int
\tilde{\rho }_h(p) dp$, we have formally, \cite{dunkel} Eq.~(9):
\begin{equation}
{\cal{S}}_h = - \int  \rho \ln \rho dx - \int \tilde{\rho }_h \ln
\tilde{\rho }_h\, dp  - \ln h = S^x + S^p  - \ln h   \label{deco}
\end{equation}
to be compared with Eq.~(\ref{unit}). The formal use of the
logarithm properties before executing integrations in the $\int
\tilde{\rho }_h \ln (h \tilde{\rho }_h)\, dp$, has left us with
the previously mentioned  issue (Section  1.2) of "literally
taking the logarithm of a dimensional argument" i. e.  that of
$\ln h$.

We recall that ${\cal{S}}_h $ is a dimensionless quantity, while
if $x$ has dimensions of length,     then the probability density
has dimensions of inverse length and  analogously in connection
with momentum dimensions.

Let us denote $x\doteq   r \delta x$ and $p\doteq \tilde{r} \delta
p$ where labels $r$ and $\tilde{r}$ are dimensionless, while
$\delta x$ and $\delta p$ stand  for respective position and
momentum  dimensional (hitherto - resolution) units. Then:
\begin{equation}
 -\int \rho  \ln
\rho dx  - \ln (\delta x)  \doteq  -\int \rho \ln (\delta x \rho
)dx  \label{34}
\end{equation}
 is a dimensionless quantity. Analogously
\begin{equation}
-\int \tilde{\rho }_h \ln \tilde{\rho }_h\, dp - \ln \delta p
\doteq -\int \tilde{\rho }_h \ln (\delta p \tilde{\rho }_h)\, dp
\end{equation}
is  dimensionless. First left-hand-side terms in two above
equations we recognize as $S^x$ and $S^p$ respectively.

Hence, formally we get a manifestly dimensionless decomposition
\begin{equation}
{\cal{S}}_h = -\int \rho \ln (\delta x \rho )dx  -\int \tilde{\rho
 }_h \ln (\delta p \tilde{\rho }_h)\, dp  + \ln {\frac{\delta x
\delta p}h} \doteq S^x_{\delta x} + S^p_{\delta p} +\ln
{\frac{\delta x \delta p}h} \label{final}
\end{equation}
instead of the previous one, Eq.~(\ref{deco}). The last identity
Eq.~(\ref{final})  gives an unambiguous meaning to the preceding
formal  manipulations with  dimensional quantities.

 As a byproduct of our discussion,
we have resolved the case of the spatially interpreted real axis,
when $x$ has dimensions of length, c.f. also \cite{ohya}:
$S^x_{\delta x}$ is the pertinent dimensionless differential
entropy definition for spatial probability densities.

{\bf Example 1:} Let us go through an explicit example involving
the Gauss density
\begin{equation}
 \rho (x)= (1/\sigma \sqrt{2\pi })\, \exp [-
(x-x_0)^2/2\sigma ^2]   \label{gaussian}
\end{equation}
 where $\sigma $ is the standard deviation
(its square stands for the variance). There holds ${\cal{S}}(\rho
) =  {\frac{1}{2}} \ln \,(2\pi e \sigma ^2)$ which is  a
dimensionless outcome.  If we pass to $x$ with dimensions of
length, then  inevitably $\sigma $ must have dimensions of length.
It is instructive to check that in this dimensional case we have a
correct dimensionless result:
\begin{equation}
S^x_{\delta x} = {\frac{1}{2}} \ln \,[2\pi e \left({\frac{\sigma
}{\delta x}}\right)^2]
\end{equation}
to be compared with Eq.~(\ref{34}).  Then we realize that
$S^x_{\delta x}$ vanishes if $\sigma /\delta x = (2\pi e )^{-1/2}
$, hence at the dimensional value  of the standard deviation
$\sigma = (2\pi e )^{-1/2} \delta x$, compare e.g. \cite{ohya}.

{\bf Example 2:} Let us invoke the
 simplest (naive) text-book  version of the Boltzmann   H-theorem, valid in case of the   rarified gas
  (of mass $m$ particles), without external forces, close to its thermal equilibrium,  under an
assumption of its  space homogeneity, \cite{huang,cercignani}.
  If the probability density function $f(v)$ is a solution of the corresponding  Boltzmann
  kinetic equation, then the Boltzmann $H$-functional  (which is
  simply the negative of the differential entropy)    $H(t) = \int f(v) \ln f(v) dv $  does not increase:
  ${\frac{d}{dt}} H(t) \leq 0 $.
 In the present case  we know that there exists an
  invariant (asymptotic) density, which in one-dimensional case has the form  $ f_*(v)
  = (m/2\pi k_BT)^{1/2}  \exp[- m (v-v_0)^2/2k_BT]$.    $H(t)$  is  known to be
   time-independent  only if $f\doteq f_*(v)$.
   We can straightforwardly, albeit formally,  evaluate $H_* = \int f_* \ln f_* dv = - (1/2)
   \ln (2\pi e k_BT/m)$  and become  faced with  an apparent dimensional difficulty, \cite{bril},
   since an argument of the logarithm is not  dimensionless.
For sure, a   consistent     integration outcome for $H(t)$ should
involve a dimensionless argument  $k_BT/m[v]^2$ instead of
$k_BT/m$,     provided  $[v]$ stands for \it  any \rm unit of
velocity; examples are $[v]= 1\, m/s$     (here $m$ stands for the
SI length unit, and not  for a mass parameter) or $10^{-5}\, m/s$.
To this end, in conformity with our previous discussion,  it
suffices to redefine
    $H_*$ as follows, \cite{bril,ohya}:
\begin{equation}
    H_* \rightarrow H_*^{[v]}=   \int f_* \ln ([v]\cdot  f_*) dv \, .
\end{equation}
Multiplying $f_*$ by $[v]$ we arrive at the dimensionless argument
of the logarithm in the above and cure the dimensional obstacle.

{\bf Remark 7:} Let us  also   mention an  effect  of the scaling
transformation upon the differential entropy, \cite{shannon}. We
denote
\begin{equation}
\rho _{\alpha, \beta }
 = \beta \, \rho  [\beta (x-\alpha )]
\end{equation}
  where $\alpha >0, \beta >0$.
  The respective Shannon entropy reads:
\begin{equation}
   {\cal{S}}(\rho _{\alpha ,\beta })  = {\cal{S}}(\rho )  -
 \ln \beta   \, .
 \end{equation}
In case of the  Gaussian $\rho $,  Eq.~(\ref{gaussian}),  we get
${\cal{S}}(\rho _{\alpha ,\beta })=  \ln [(\sigma /\beta )
\sqrt{2\pi e}]$.
 An obvious interpretation is
 that the $\beta $-scaling transformation of $\rho (x-\alpha ) $ would  broaden this density if
 $\beta <1$ and would  shrink when $\beta >1$. Clearly, ${\cal{S}}(\rho _{\alpha ,\beta
 })$ takes the value $0$ at $\sigma = (2\pi e )^{-1/2} \beta $ in analogy with our previous
 dimensional considerations. If an  argument of $\rho $ is assumed
 to have dimensions, then the scaling transformation with the
 dimensional $\beta $ may be interpreted as a method to restore
 the dimensionless differential entropy value.

 \section{Localization: differential entropy and Fisher
information}

We  recall  that among \it all \rm  one-dimensional distribution
functions  $\rho (x)$   with a finite   mean,  subject to the
constraint  that the standard deviation is fixed at $\sigma $, it
is the Gauss function  with half-width $\sigma $ which sets  a
maximum  of the
 differential entropy, \cite{shannon}.  For the record, let us add that if
 only the mean is  given for  probability density functions on $R$, then there is
  no maximum entropy distribution in their set.

Let us     consider  the Gaussian probability density on the real
line $R$ as a reference  density function: $ \rho (x)= (1/\sigma
\sqrt{2\pi })\, \exp [- (x-x_0)^2/2\sigma ^2]$, in conformity with
Eq.~(\ref{gauss1}), but without any restriction on the value of
$x\in R$.

 The differential entropy of the Gauss density  has  a simple analytic form, independent of the mean
value $x_0$ and  maximizes an inequality:
\begin{equation}
 {\cal{S}}(\rho ) \leq  {\frac{1}{2}} \ln \,(2\pi e \sigma ^2) \, . \label{bound}
\end{equation}
This imposes a useful bound upon the  entropy power:
\begin{equation}
{\frac{1}{\sqrt{2\pi e}}} \exp [{\cal{S}}(\rho )] \leq  \sigma
\label{power}
\end{equation}
with an obvious bearing on the spatial localization  of the
density $\rho $,
 hence spatial (un)certainty of position measurements.   We can say that almost surely,  with
 probability  $0.998$,  there is   nothing  to be found (measured) beyond the interval of the
length $6\sigma $ which is centered about the mean value $x_0$  of
the Gaussian density
 $\rho $.

 Knowing that for arbitrary density functions  the differential entropy
 Eq.~(\ref{two}) is unbounded form below  and from above, we realize that in the  subset of  all
  densities with a finite mean and  a fixed variance $\sigma ^2$, we  actually have an upper bound
  set  by Eq.~(\ref{bound}).
However, in contrast to  coarse grained entropies which are always
nonnegative,   even for relatively large   mean deviation  $\sigma
< 1/\sqrt{2\pi e} \simeq  0.26 $ the  differential entropy
${\cal{S}}(\rho )$   is  negative.

Therefore, quite apart from the previously discussed  direct
information theory links, c.f. Eqs.~(\ref{grating}),  (\ref{ineq})
and (\ref{appr}), the major role of the differential entropy is to
be a measure of   localization in the "state space" (actually,
configuration space)
 of the system, \cite{hall,pipek,pipek1}.

Let us consider a one-parameter family of probability densities
$\rho _{\alpha }(x)$
  on $R$ whose first (mean) and second  moments (effectively, the variance) are finite.
The parameter-dependence is here not completely arbitrary and we
assume standard regularity properties that allow to differentiate
various functions  of $\rho _{\alpha }$ with respect to  the
parameter
 $\alpha $ under the sign of an  (improper) integral, \cite{cramer}.

Namely, let us denote  $ \int x\rho _{\alpha }(x) dx = f(\alpha )$
and $\int x^2\rho _{\alpha } dx < \infty $. We demand that as a
function of $x\in R$, the  modulus of the partial derivative
$\partial \rho _{\alpha }/ \partial \alpha $ is  bounded by a
function $G(x)$ which together with  $x G(x)$ is integrable on
$R$.  This implies,  \cite{cramer}, the existence of $\partial f
/\partial \alpha $ and an important inequality:
\begin{equation}
\int (x- \alpha )^2 \rho _{\alpha } dx \cdot \int \left(
 {\frac{\partial ln \rho _{\alpha }}{\partial \alpha }}
 \right)^2 \rho _{\alpha } dx  \label{cramer}
 \geq
 \left( \frac{df(\alpha )}{d\alpha }\right)^2
\end{equation}
directly resulting from
\begin{equation}
{\frac{d f}{d\alpha }} = \int [(x - \alpha ) \rho _{\alpha
}^{1/2}] [{\frac{\partial (\ln \rho _{\alpha })} {\partial \alpha
}} \rho _{\alpha }^{1/2}] dx
\end{equation}
via the standard Schwarz inequality. An  equality  appears  in
Eq.~({\ref{cramer}) if $\rho _{\alpha }(x)$ is the  Gauss function
with
 mean value $\alpha $.

At this point let assume that the mean value of $\rho _{\alpha }$
actually
  equals $\alpha $ and we fix at  $\sigma ^2$ the  value
   $\langle (x-\alpha )^2\rangle = \langle x^2 \rangle -  \alpha ^2 $
of  the variance  (in fact, standard deviation from the mean value
) of the probability density $\rho _{\alpha }$. The previous
inequality Eq.~(\ref{cramer})  now  takes the  familiar  form:
\begin{equation}
 {\cal{F}}_{\alpha } \doteq  \int   {\frac{1}{\rho _{\alpha }}} \left({\frac{\partial
 \rho _{\alpha}}{\partial \alpha }}
\right)^2\, dx \geq {\frac{1}{\sigma ^2}} \label{Fisher}
\end{equation}
where  an integral on the left-hand-side is the so-called Fisher
information of $\rho _{\alpha }$,
 known to   appear  in various problems of   statistical estimation
 theory, as well as an ingredient of a number of   information-theoretic inequalities,
 \cite{cramer,stam,dembo,hall,hall1}.  In view  of  ${\cal{F}}_{\alpha } \geq 1/\sigma ^2$,
 we realize that the Fisher  information  is   more sensitive indicator of the wave packet
  localization   than the entropy power, Eq.~(\ref{power}).

Let us  define $\rho _{\alpha }(x) \doteq \rho (x-\alpha )$. Then,
the Fisher information  ${\cal{F}}_{\alpha }\doteq  {\cal{F}}$ is
no longer  the mean value $\alpha $-dependent   and  can be
readily transformed to the conspicuously quantum mechanical form
(up to a factor $D^2$ with $D=\hbar /2m$):
\begin{equation}
{\frac{1}{2}} {\cal{F}} =   {\frac{1}{2}} \int {\frac{1}{\rho }}
\left({\frac{\partial \rho }{\partial x }} \right)^2\, dx  = \int
\rho \cdot  {\frac{u^2}{2}} dx = - \langle Q \rangle
\label{Fisher1}
  \end{equation}
 where  $u\doteq  \nabla \ln \rho $   is named an osmotic velocity field, \cite{nelson,gar}, and
an average $\langle Q\rangle =\int  \rho \cdot Q dx$ is carried
out with respect to the function
 \begin{equation}
Q= 2 {\frac{\Delta \rho ^{1/2}}{\rho ^{1/2}}}\, .
\label{potential}
\end{equation}
As a consequence of Eq.~(\ref{Fisher}), we have $- \langle
Q\rangle \geq 1/2\sigma ^2$ for all relevant probability densities
with  \it  any \rm finite mean and  variance fixed at $\sigma ^2$.

When multiplied by $D^2$,   the above  expression  for $Q(x)$
notoriously appears in the hydrodynamical formalism    of quantum
mechanics  as the  so-called  de Broglie-Bohm quantum potential
($D=\hbar /2m$). It appears as well  in   the corresponding
formalism  for diffusion-type  processes, including the standard
Brownian motion (then,  $D=k_BT/m\beta $, see e.g.
\cite{gar,gar1,gar2}.

An important inequality, valid under an assumption  $\rho _{\alpha
}(x) = \rho (x- \alpha )$, has been proved in \cite{stam}, see
also \cite{dembo,carlen1}:
\begin{equation}
{\frac{1}{\sigma ^2}} \leq    (2\pi e) \, \exp[- 2 {\cal{S}}(\rho
)] \leq   {\cal{F}} \label{first}
\end{equation}
It tells  us that the lower bound for the Fisher information is in
fact  given a  sharper  form  by means of the (squared) inverse
entropy power. Our  two information measures  appear to be
correlated.\\

{\bf Remark 8:} Let us point out that the Fisher information
${\cal{F}}(\rho )$ may blow up to infinity under a number of
circumstances, \cite{hall1}: when $\rho $ approaches the Dirac
delta behavior, if $\rho $    vanishes over some  interval in $R$
or   is discontinuous. We observe that  ${\cal{F}}>0$ because  it
may   vanish only when  $\rho $ is constant  everywhere on $R$,
hence  when $\rho $ is \it not  \rm a probability density.\\

{\bf Remark 9:}
   The values  of ${\cal{F}}(\rho _{\alpha })$  and
${\cal{S}}(\rho _{\alpha })$ are $\alpha $-independent  if we
consider  $\rho _{\alpha }(x)=\rho (x- \alpha )$. This reflects
the translational invariance of the Fisher and Shannon information
measures, \cite{frieden}.  The scaling transformation  $\rho
_{\alpha, \beta }
 = \beta \, \rho  [\beta (x-\alpha )]$, where $\alpha >0, \beta >0$,
   transforms  Eq.~(\ref{power})  to  the form    $(2\pi e)^{-1/2}\,
  \exp [{\cal{S}}(\rho _{\alpha ,\beta })]  \leq     \sigma /\beta
  $.\\

Under an additional decomposition/factorization  ansatz (of the
quantum mechanical  $L^2(R^n)$
 provenance) that $\rho (x) \doteq |\psi |^2(x)$,     where a real or complex
 function  $\psi  =\sqrt{\rho } \exp(i\phi )$ is a normalized  element of
$L^2(R)$, another important inequality holds true,
\cite{stam,petz}:
\begin{equation}
{\cal{F}}  =  4 \int \left({\frac{\partial \sqrt{\rho }}{\partial
x}} \right)^2 dx \leq   16 \pi ^2 \tilde{\sigma }^2 \, ,
\label{second}
\end{equation}
provided  the Fisher information takes finite values. Here,
$\tilde{\sigma }^2$ is  the variance of the "quantum mechanical
momentum canonically conjugate to the position observable", up to
(skipped) dimensional factors. In the above, we have exploited the
Fourier transform $\tilde{\psi } \doteq ({\cal{F}} \psi)$  of
$\psi $ to arrive at $\tilde{\rho }\doteq |\tilde{\psi }|^2$ of
Eq.~(\ref{one})  whose variance the above $\tilde{\sigma }^2$
actually is.

In view of  two previous inequalities (\ref{first}),
(\ref{second})  we find  that not only the Fisher information, but
also  an entropy power is bounded from below and  above. Namely,
we have:
\begin{equation}
{\frac{1}{\sigma ^2}} \leq {\cal{F}} \leq  16\pi ^2 \tilde{\sigma
}^2 \label{in1}
\end{equation}
which implies $1/2\sigma ^2 \leq  -\langle Q  \rangle \leq 8\pi ^2
\tilde{\sigma }^2 $  and furthermore
\begin{equation}
{\frac{1}{4\pi \tilde{\sigma }}} \leq {\frac{1}{\sqrt{2\pi e}}}\,
\exp[{\cal{S}}(\rho)]  \leq  \sigma \, . \label{in2}
\end{equation}
 as a complement to Eq.~(\ref{power}).  Most important outcome of Eq.~({\ref{in2}) is that the differential
 entropy ${\cal{S}}(\rho) $  typically  may be expected to be a well behaved   quantity:   with finite
  both  lower and upper bounds.

 We find rather interesting that the Heisenberg indeterminacy  relationship  Eq.~(\ref{heisenberg}),
 which is   normally interpreted to set  a lower bound on the
experimentally  accessible \it  phase-space \rm  data (e.g.
volume), according to Eq.~(\ref{in2})
  ultimately  had appeared to  give rise to     lower and upper bounds upon  the
  \it  configurational \rm (spatial)    information measure   and thence -  the   uncertainty
  (information) measure.

To our knowledge, the  inequalities Eqs.~(\ref{in1}) and
(\ref{in2}), although implicit in various  information theory
papers, see especially \cite{stam} and  \cite{petz},  hitherto
were never explicitly spelled out.

\section{Asymptotic approach towards equilibrium:  Smoluchowski processes}

\subsection{Random walk}

Let us consider a classic example of a one-dimensional random walk
where a particle is assumed to  be displaced  along $R^1$ with a
probability $1/2$ forth and back, each step being of a unit
length, \cite{chandra}. If one begins from  the origin $0$, after
$G$ steps a particle can found at any of the  points
$-G,-G+1,...-1,0,1,...,G$.  The probability that after $G$
displacements a particle can be found at the point  $g\in [-G,G]$
is given by the Bernoulli distribution:
\begin{equation}
P_g = {\frac{G!}{[(1/2)(G+g)]![(1/2)(G-g)]!}}
\left({\frac{1}2}\right)^G   \label{walk}
\end{equation}
where $\sum_{g=-G}^{G}  P_g = 1$.

We are interested in the asymptotic formula, valid for large $G$
and $g \ll G$. (Note that even  for relatively  small value of
$G=20$, and $|g|\leq 16$, an accuracy level  is satisfactory.)
There holds:
\begin{equation}
\ln P_g  \simeq   - {\frac{g^2}{2G}}  + {\frac{1}2} \ln
\left({\frac{2}{\pi G}}\right)
\end{equation}
and, accordingly
\begin{equation}
P_g \simeq  \left({\frac{2}{\pi G}}\right)^{1/2} \exp \left(-
{\frac{g^2}{2G}}\right) \, .
\end{equation}

We assume  $r\sim 10^{-6}m$,  to be  a  grating unit (i.e. minimal
step length for a walk).
 Let $r\ll \Delta x \ll G$ (size $\Delta x \sim  10^{-4}m$ is quite satisfactory).
 For large $G$ and $|g| \ll G$, we denote $x\doteq g\cdot r$ and  ask for a probability $\rho _G(x) \Delta x$
  that a particle can be found in the interval $[x,x+\Delta x]$ after $G$ displacements. The result is
  \cite{chandra}:
\begin{equation}
  \rho (x) = {\frac{1}{2r}}   P_g = {\frac{1}{(2\pi Gr^2)^{1/2}}} \exp \left(- {\frac{x^2}{2Gr^2}}
  \right) \label{heat}
 \end{equation}
and  by assuming that a particle suffers $k$ displacements per
unit time, we  can give Eq.~(\ref{heat}) the familiar form of the
heat kernel:
\begin{equation}
\rho (x,t) = {\frac{1}{(4\pi Dt)^{1/2}}} \exp \left( -
{\frac{x^2}{4Dt}}\right)  \label{kernel}
\end{equation}
with the diffusion coefficient $D=kr^2/2$. It is a fundamental
solution of the heat equation
 $\partial _t \rho = D \Delta \rho $ which is the Fokker-Planck equation for the Wiener  process.

The differential entropy of the above  time-dependent density
  reads:
\begin{equation}
{\cal{S}}(t) =  (1/2) \ln (4\pi e Dt) \label{heatkernel}
\end{equation}
 and its time evolution clearly displays the
  localization  uncertainty growth.
  By means of the formula Eq.~(\ref{appr})    we can quantify the differential  entropy dynamics for
  all solutions of the heat equation.

Since the heat kernel determines  the transition probability
density    for the Wiener process (free Brownian motion in $R$),
 by setting $x\rightarrow x-x'$ and $t \rightarrow t-t' \geq 0 $,  we can  replace the previous
 $\rho (x,t)$ of Eq.~(\ref{kernel}) by    $p(x-x',t-t')$. This  transition density  allows  to deduce
  \it any \rm  given solution $\rho (x,t)$ of the heat  equation from its past data, according to:
 $\rho (x,t) = \int p(x-x',t-t') \rho  (x',t,) dx'$.
  In particular, we can consider the process starting at $t'=0$ with any initial density $\rho _0(x)$.

Let $\rho _{\upsilon }$ denote  a convolution of a probability
density  $\rho $
 with  a Gaussian
probability density having variance $\upsilon $.   The transition
density of the  Wiener process generates such a convolution for
$\rho _0$,
 with $\upsilon = \sigma ^2
\doteq 2Dt$.   Then,
 de Bruijn identity,  \cite{stam,hall},
 $d{\cal{S}}(\rho _{\upsilon })/d\upsilon =  (1/2) {\cal{F}}(\rho _{\upsilon })$,
 directly  yields   the  information entropy time rate  for ${\cal{S}}(\rho )= {\cal{S}}(t)$:
    \begin{equation}
{\frac{d{\cal{S}}}{dt}} =  D \cdot{\cal{F}}  =    D\cdot \int
{\frac{(\nabla  \rho )^2}{\rho }} dx > 0  \label{rate}\, .
\end{equation}
The Fisher information ${\cal{F}}(\rho )$ is the $\alpha =0$
version of the  general  definition given in  Eqs.~(\ref{Fisher})
and (\ref{Fisher1}).  The derivation of
  Eq.~(\ref{rate})  amounts to differentiating an
 $\upsilon $-dependent   integrand   under the sign of an improper integral, \cite{cramer,hall1})

 The monotonic  growth of ${\cal{S}}(t)$  is paralleled by   linear in time  growth of
$\sigma (t)$  and the decay of  ${\cal{F}}$, hence quantifies  the
uncertainty (disorder) increase related to  the "flattening" down
of $\rho $, see also \cite{hall1,frieden}.

\subsection{Kullback entropy versus differential entropy}

We  emphasize that  in the present paper  we  have deliberately
avoided the use of the relative  Kullback-Leibler entropy,
\cite{kullback,sobczyk,mackey}.  This entropy  notion  is often
invoked to tell "how far from each other"  two probability
qdensities are. The  Kullback entropy is particularly useful if
one
 investigates an approach of the system toward (or its deviation
from) equilibrium, this being normally represented by a stationary
density function, \cite{lasota,risken}. In this context, it is
employed to investigate a major issue of the   dynamical origins
of the increasing  entropy, see \cite{mackey,lasota,tyran}.
Consult also both  standard motivations and apparent  problems
encountered in connection with the celebrated Boltzmann's
$H$-theorem, \cite{huang,cercignani} and \cite{qian}.

However,  a  reliability  of the Kullback entropy   may be
questioned in case of general parameter-dependent densities.  In
particular, this entropy  fails to quantify properly  certain
features of  a non-stationary dynamics  of probability densities.
Specifically if we  wish to  make a "comparison" of once given
density function to itself, but  at different stages (instants) of
its time evolution.

Let us consider  a one parameter family
 of  Gaussian densities   $\rho _{\alpha  } = \rho (x-\alpha  )$, with the mean  $\alpha  \in R$ and
 the  standard deviation fixed at $\sigma $.  These densities  are not differentiated
 by the information  (differential)  entropy and share   its  very same value  $ {\cal{S}}_{\sigma }
 =   {\frac{1}{2}} \ln \,(2\pi e \sigma ^2)$  independent of $\alpha $.

If we admit $\sigma $ to be another free parameter, a
two-parameter family of Gaussian densities
  $\rho _{\alpha  } \rightarrow \rho _{\alpha , \sigma }(x)$ appears.  Such
  densities,  corresponding to   different  values of  $\sigma $ and $\sigma '$ do
admit an "absolute comparison" in terms of the Shannon entropy, in
accordance with Eq.~(\ref{appr}):
\begin{equation}
{\cal{S}}_{\sigma '} - {\cal{S}}_{\sigma } =  \ln \, \left(
\frac{\sigma '}{{\sigma }}\right) \, .  \label{comparison}
\end{equation}

By denoting    $\sigma \doteq \sigma (t) = \sqrt{2Dt}$ and $\sigma
' \doteq \sigma (t')$ we  make  the non-stationary (heat kernel)
density  amenable to the "absolute comparison"  formula
 at different time instants $t'>t >0$: $(\sigma '/ \sigma ) = \sqrt{t'/t}$.

In the above we have "compared"  differential entropies of quite
akin, albeit different,  probability densities. Among many
inequivalent ways to evaluate the "divergence" between probability
distributions, the relative (Kullback) entropy is typically used
to quantify such   divergence  from the a priori  prescribed
reference density, \cite{risken,lasota}.

We define the  Kullback  entropy ${\cal{K}}(\theta ,\theta')$ for
a one-parameter family of probability densities $\rho _{\theta}$,
so that the   "distance" between  any two densities in this family
can be directly  evaluated.  Let  $\rho _{\theta '}$ stands for
the prescribed  (reference) probability density. We have,
\cite{kullback,risken,sobczyk}:
\begin{equation}
{\cal{K}}(\theta ,\theta ') \doteq  {\cal{K}}(\rho _{\theta }|\rho
_{\theta '}) =    \int \rho _{\theta }(x)\, \ln {\frac{\rho
_{\theta }(x)}{\rho _{\theta '}(x)}}\, dx \, .
\end{equation}
which, in view of the concavity of the function $f(w) = - w \ln
w$,  is  positive.

 Let us indicate that the negative of
${\cal{K}}$, ${\cal{H}}_c \doteq - {\cal{K}}$,    named the
conditional entropy \cite{sobczyk}, is predominantly used in the
literature \cite{sobczyk,tyran,lasota}  because of  its  affinity
(regarded as a  formal  generalization)  to the differential
entropy. Then e.g. one investigates an approach of $-{\cal{K}}$
towards its maximum (usually achieved at the value zero) when  a
running density is bound to have a  unique stationary asymptotic,
\cite{tyran}.

If we take  $\theta ' \doteq \theta + \Delta \theta$ with $ \Delta
\theta  \ll 1$,  the following approximate formula holds true
under a number of  standard assumptions, \cite{kullback}:
\begin{equation}
{\cal{K}}(\theta ,\theta + \Delta  \theta ) \simeq {\frac{1}{2}}
{\cal{F}}_{\theta }\, \cdot  (\Delta \theta )^2  \label{approx}
\end{equation}
where ${\cal{F}}_{\theta }$  denotes  the Fisher information
measure, previously  defined by Eq.~(\ref{Fisher}).  With this
proviso, we can  evaluate the Kullback distance within
 a two-parameter $(\alpha , \sigma )$ family of Gaussian
densities, by taking $\theta \rightarrow \alpha $.

Passing to $\alpha  '= \alpha  +\Delta \alpha  $ at  a fixed value
of  $ \sigma $  we arrive at:
\begin{equation}
{\cal{K}}(\alpha ,\alpha  + \Delta  \alpha  ) \simeq
{\frac{(\Delta \alpha )^2} {2\sigma ^2}}\, .
\end{equation}
For the record, we  note   that  the  respective Shannon entropies
do  coincide: ${\cal{S}}_{\alpha } = {\cal{S}} _{\alpha + \Delta
\alpha }$.

Analogously, we can  proceed with respect to the label $\sigma $
at  $\alpha $ fixed:
\begin{equation}
{\cal{K}}(\sigma , \sigma + \Delta \sigma ) \simeq {\frac{(\Delta
\sigma )^2}{ \sigma ^2}}
\end{equation}
when, irrespective of  $\alpha $:
\begin{equation}
{\cal{S}}_{\sigma + \Delta \sigma } - {\cal{S}}_{\sigma } \simeq
{\frac{\Delta \sigma } {\sigma }}\, .
\end{equation}

By choosing  $\theta \rightarrow \sigma ^2$ at  $\alpha $ fixed,
we get (now the variance $\sigma ^2$ is modified by its increment
$\Delta (\sigma ^2)$):
\begin{equation}
{\cal{K}}(\sigma ^2 , \sigma ^2 + \Delta (\sigma ^2)) \simeq
{\frac{[\Delta (\sigma ^2)]^2}{4 \sigma ^4}}
\end{equation}
while
\begin{equation}
{\cal{S}}_{\sigma ^2 + \Delta (\sigma ^2) } - {\cal{S}}_{\sigma
^2} \simeq {\frac{\Delta (\sigma  ^2)}{2\sigma ^2 }}
\end{equation}
which, upon  identifications  $\sigma ^2 = 2Dt$ and $\Delta
(\sigma ^2)= 2D\Delta t$, sets an obvious connection with the
 differential $(\Delta {\cal{S}})(t)$ and thence   with  the
  time derivative $\dot{\cal{S}} = 1/2t$ of the heat kernel
differential entropy, Eq.~(\ref{heatkernel}) and the de Bruijn
identity.

Our  previous observations are a special case of more general
reasoning. Namely, if we consider  a two-parameter $\theta \doteq
(\theta _1,\theta _2)$  family of densities, then instead of
Eq.~(\ref{approx}) we would have  arrived at
\begin{equation}
{\cal{K}}(\theta ,\theta + \Delta  \theta ) \simeq {\frac{1}{2}}
 \sum_{i,j} {\cal{F}}_{ij}\, \cdot  \Delta \theta _i  \Delta \theta _j
\end{equation}
where $i,j, = 1,2$ and the Fisher information matrix
${\cal{F}}_{ij}$ has the form
\begin{equation}
{\cal{F}}_{ij} = \int \rho _{\theta } {\frac{\partial \ln \rho
_{\theta } }{\partial \theta _i}}  \cdot {\frac{\partial \ln \rho
_{\theta } }{\partial \theta _j}}\, dx \, .
\end{equation}

In case of Gaussian densities, labelled  by  independent
parameters $\theta _1 = \alpha $ and $\theta _2 = \sigma $
(alternatively $\theta _2 = \sigma ^2$), the Fisher matrix is
diagonal and defined in terms of previous entries
${\cal{F}}_{\alpha }$ and ${\cal{F}}_{\sigma }$ (or
${\cal{F}}_{\sigma ^2}$).

It is useful  to  note  (c.f. also \cite{tyran})  that in
self-explanatory notation, for  two $\theta $ and $\theta '$
Gaussian densities there holds:
\begin{equation}
{\cal{K}}(\theta ,\theta ')=  \ln {\frac{\sigma '}{\sigma }} +
{\frac{1}{2}}({\frac{\sigma ^2}{{\sigma '}^2}} - 1) +
{\frac{1}{2{\sigma '}^2}} (\alpha - \alpha ')^2  \label{kulgau}
\end{equation}
The first entry in Eq.~(\ref{kulgau}) coincides with the
"absolute comparison formula" for Shannon entropies,
Eq.~(\ref{comparison}). However for $|\theta ' - \theta |\ll 1$,
hence in the regime of interest for us,  the second term dominates
the first one.

Indeed, let us set $\alpha ' = \alpha $ and consider $\sigma ^2 =
2Dt$, $\Delta (\sigma ^2)= 2D\Delta t$. Then ${\cal{S}}(\sigma ')
- {\cal{S}}(\sigma ) \simeq \Delta t/2t$, while ${\cal{K}}(\theta
,\theta ') \simeq (\Delta t)^2/ 4t^2$. Although, for finite
increments $\Delta t$ we have
\begin{equation}
 {\cal{S}}(\sigma ') -
{\cal{S}}(\sigma )\simeq \sqrt{ {\cal{K}}(\theta ,\theta ')}\simeq
{\frac{\Delta t}{2t}} \, ,
\end{equation}
the time derivative
 notion $\dot{\cal{S}}$  can be defined exclusively   for the differential entropy,
 and is meaningless  in terms of  the Kullback "distance".

 Let us mention that no such obstacles arise  in the standard cautious  use of the
 relative  Kullback entropy ${\cal{H}}_c$.  Indeed, normally one
 of the involved densities stands for the  stationary  reference one
 $\rho _{\theta '}(x) \doteq \rho _*(x)$,  while  another evolves in time
 $\rho _{\theta }(x) \doteq \rho (x,t)$, $t\in R^+$, thence
 ${\cal{H}}_c(t) \doteq - {\cal{K}}(\rho _t|\rho _*)$,   see e.g.  \cite{lasota,tyran}.

\subsection{Entropy dynamics in the  Smoluchowski process}

We  consider spatial Markov diffusion processes in $R$  with a
diffusion coefficient  (constant or time-dependent) $D$ and admit
them to drive space-time inhomogeneous probability densities $\rho
= \rho (x,t)$. In the previous section we have addressed the
special case of the   free  Brownian motion characterized by  the
current velocity  (field)  $v \doteq  v(x,t) = - u(x,t) =
 D \nabla \ln \rho  (x,t)$  and the diffusion current
 $j \doteq v \cdot \rho $ which obeys the continuity equation
  $\partial _t\rho = - \nabla j$, this in turn being equivalent to the
  heat equation.

It is instructive to notice that  the   gradient of a
potential-type function
 $Q = Q(x,t)$,  c.f. Eq.~(\ref{potential}),
 entirely composed in terms of $u$:
\begin{equation}
Q=2D^{2}{\frac{\Delta \rho ^{1/2}} {\rho ^{1/2}}} = {\frac{1}2}
u^2 + D \nabla \cdot u  \label{potential1}
\end{equation}
  almost trivially  appears (i.e. merely as a consequence of the heat equation, \cite{gar,czopnik})
  in the   hydrodynamical (momentum) conservation law  appropriate for the free Brownian motion:
\begin{equation}
\partial _t {v} + (v \cdot
{\nabla }) {v} = - {\nabla }Q \, . \label{free}
\end{equation}

 A straightforward  generalization refers   to  a diffusive dynamics of a
 mass $m$ particle in the external field of force, here taken to
 be conservative: $F=F(x) =-{\nabla } V$.
 The associated  Smoluchowski diffusion process
with a forward drift ${b}(x) = \frac{F}{m\beta }$
 is analyzed in terms of the   Fokker-Planck equation for the spatial probability
density $\rho(x,t)$, \cite{risken,hasegawa,rubi,kurchan}}:
\begin{equation}
\partial _t\rho =
D\triangle \rho -  {\nabla } (b\cdot  \rho )\,  \label{Fokker}
\end{equation}
with the initial data $\rho _0(x) = \rho (x,0)$.

Note that if things are specialized to the \it standard \rm
Brownian motion in  an external force field, we know  a priori
(due to the Einstein fluctuation-dissipation relationship,
\cite{chandra}) that $D=\frac{k_BT}{m\beta }$, where  $\beta $ is
interpreted as  the friction (damping) parameter,  $T$ is the
temperature of the bath, $k_B$ being the Boltzmann constant .

We assume, modulo restrictions upon drift function
\cite{eberle,qian1}, to resolve the Smoluchowski dynamics in terms
of (possibly non-unique) Markovian diffusion-type processes. Then,
the following  compatibility equations follow in  the form of
hydrodynamical   conservation laws for the diffusion process,
\cite{gar,czopnik}:
\begin{eqnarray}
\partial _{t}\rho + {\nabla }(
{v} \rho )%
 &=&0 \\
(\partial _{t} + {v}
\cdot {\nabla }) {v}%
 &=& {\nabla } ( \Omega -Q)\,  \label{law}
\end{eqnarray}
where, not to confuse this notion with the previous force field
potential  $V$,  we denote by $\Omega  (x)$
  the so-called volume potential for the process:
\begin{equation}
\Omega =\frac{1}{2}\left( \frac{F}{\ m\beta }\right) ^{2}+D{\nabla
}\cdot \left( \frac{F}{\ m\beta }\right) \, ,  \label{Omega}
\end{equation}
where the functional form of  $Q$  is given by
Eq.~(\ref{potential1}). Obviously the free Brownian  law,  Eq.
(\ref{free}),  comes out as the  special case.

In the above (we use  a short-hand notation $v \doteq v(x,t)$):
\begin{equation}
v \doteq   b - u
 = \frac{%
 F}{\ m\beta }- D\frac{\nabla
\rho }{\rho }
\end{equation}
 defines  the  current velocity of Brownian particles in external force field.
 This formula allows us  to   transform the continuity equation into the
 Fokker-Planck equation  and back.

 With a solution  $\rho (x,t)$ of  the Fokker-Planck equation, we
  associate a differential (information) entropy \rm  ${\cal{S}}(t)=  - \int \rho \, \ln \rho \,  dx$
which is typically  \it not \rm a conserved quantity, \cite{
nicolis}-\cite{halliwell}. The  rate of change in  time  of
${\cal{S}}(t)$   readily follows.

Boundary restrictions  upon  $\rho $, $v \rho $ and $b \rho $ to
vanish at spatial infinities (or  at finite spatial volume
boundaries) yield the  rate  equation:
\begin{equation}
{\frac{d{\cal{S}}}{dt}}  = \int [\rho \, (\nabla \cdot b)
 + D \cdot  {\frac{(\nabla \, \rho )^2}\rho }]\, dx  \label{force}
\end{equation}
to be compared with the previous, $b =0$ case,  Eq. (\ref{free}).

Anticipating further discussion, let us stress that  even in case
of
 plainly irreversible diffusive  dynamics, it is by no means obvious  that
  the differential entropy should grow,  decay (diminish) or show up a mixed behavior.
It is often tacitly assumed that one  should "typically"  have
$\dot{\cal{S}}>0$ which
 is not true, \cite{qian,catalan}.

 We can  rewrite  Eq. (\ref{force})  in  a number of equivalent forms,
like e.g. (note that $\langle u^2 \rangle = - D \langle \nabla
\cdot u \rangle $)
\begin{equation}
D \dot{\cal{S}} \doteq  D \left< {\nabla }\cdot
 {b} \right>  +  \left< {u}^2\right> =  D\langle \nabla \cdot
 v\rangle
 \,
 \end{equation}
and specifically, \cite{qian,qian1}, as    the  major  entropy
balance equation:
\begin{equation}
D \dot{\cal{S}}  =  \langle {v}^2\rangle
    -  \langle b\cdot {v}  \label{balance}
 \rangle =  -  \langle {v}\cdot
  {u} \rangle \,
 \end{equation}
where $\langle  \cdot  \rangle $ denotes the mean value with
respect to $\rho $.

This  balance  equation  is extremely   persuasive, since  $b=
F/( m\beta )$   and  $j = v \rho $  combine  into   a
characteristic   "power release" expression:
\begin{equation}
{\frac{d{\cal{Q}}}{dt}} \doteq  {\frac{1}D} \int {\frac{1}{m\beta
}}
 {F} \cdot  {j}\,  dx =
{\frac{1}D} \left\langle b\cdot v  \label{heat1}
 \right\rangle \, .
 \end{equation}
Like in case of not necessarily  positive $\dot{\cal{S}}$, the
"power release"  expression  $\dot{\cal{Q}}$ may be positive
which  represents the power removal to the environment,  as well
as  negative which corresponds to the power absorption from the
  environment.

In the formal thermodynamical lore, in the   above  we deal here
with the time rate at which the mechanical work per unit of mass
may possibly being dissipated  (removed to the reservoir)  in the
form of heat,
 in the course of the  Smoluchowski  diffusion process: $k_BT \dot{{\cal{Q}}}= \int F\cdot j\,  dx$,
 with  $T$ being  the temperature of the bath.
 When there is no external forces, we have $b=0$, and then the differential  entropy time   rate  formula
 for  the  free Brownian motion  Eq.~(\ref{rate})  reappears.

 On the other hand, the positive terms in Eq.~(\ref{balance}) and Eq.~(\ref{rate})   represent
   the rate at which information  entropy is  put  (pumped) into the diffusing  system  by
the  thermally active environment,  thus causing a
disorder/uncertainty growth.
    This particular "entropy  production"  rate    may  possibly  be
     counterbalanced (to this end we need external forces)
by the heat removal due to dissipation, according to:
\begin{equation}
\frac{d{\cal{S}}}{dt} = \left( {\frac{d{\cal{S}}}{dt}}\right)_{in}
- {\frac{d{\cal{Q}}}{dt}}\,  \label{balance1}
\end{equation}
where $\dot{\cal{Q}}$ is defined in Eq.~(\ref{heat1}) while
$(\dot{\cal{S}})_{in} = (1/D) \langle v^2\rangle $.\\

{\bf Remark 10:}  In Refs.~\cite{qian1,qian,qian2}  a
measure-theoretic and probabilistic justification was given  to an
interpretation of $(1/D)\langle v^2\rangle $ as the \it  entropy
production  \rm rate  of  the (originally - stationary) diffusion
process with the current velocity $v$.  We would like to point out
that traditionally, \cite{ruelle,igarashi,gaspard},  the
 statistical mechanical notion of an  entropy production refers to
the  excess entropy that  is  pumped \it out \rm  of the system.
An alternative statement tells about the
 entropy  production \it  by  \rm  the physical   system \it   into  \rm  the
thermostat. In the present discussion, an   increase  of the
information entropy of the Smoluchowski  process definitely occurs
due to  the thermal environment: the differential  entropy
 is being generated (produced) \it in \rm  the physical system  \it
by  \rm  its  environment.\\

Of particular interest  is the case of  constant information
entropy $\dot{\cal{S}} =0$ which amounts to  the existence of
steady states. In the simplest case, when the
 diffusion current  vanishes, we  encounter the primitive  realization of the state of  equilibrium
  with an invariant density $\rho $. Then,    $b=u = D \nabla  \ln \rho  $
 and we  readily  arrive at the classic  equilibrium identity  for the Smoluchowski process:
\begin{equation}
-(1/k_BT)\nabla V = \nabla \ln\, \rho  \, \label{eq}
\end{equation}
which determines the functional form of the invariant density  in
case of  a given conservative force field, \cite{qian1,risken}.
There is an ample discussion in  Ref.~\cite{qian1}
 of how these properties match with  time reversal of the stationary diffusion process and the
 vanishing  of  the entropy  production (in our lore, see e.g. \cite{gaspard,dorfman})
 rate  $(\dot{\cal{S}})_{in}$.

Coming back to the general discussion, let us define the so-called
thermodynamic force $F_{th}\doteq  v/D$  associated with the
Smoluchowski diffusion  and introduce its  corresponding
time-dependent
 potential function $\Psi (x,t)$:
 \begin{equation}
 k_BT\, F_{th}  = F - k_BT\, \nabla  \ln \rho
 \doteq - {\nabla } \Psi  \, .
  \end{equation}

Notice that   $v=  - (1/m\beta ) \nabla \Psi $. In the absence of
external forces (free Brownian motion), we obviously get $F_{th} =
- \nabla \ln \rho = - (1/D) u $.

The  mean value of the potential
\begin{equation}
\Psi = V + k_BT \ln \rho  \label{Helmholtz}
\end{equation}
 of the thermodynamic force associates  with the diffusion process  an obvious  analogue of the
   Helmholtz free energy:
\begin{equation}
\left< \Psi \right> = \left< V\right> - T\, {\cal{S}}_G
\end{equation}
where the dimensional version ${\cal{S}}_G  \doteq k_B {\cal{S}}$
of  information entropy has been introduced  (actually, it is a
direct configuration-space  analog of the Gibbs entropy). The
expectation value  of the mechanical force potential $ \left<
V\right>$ plays here the role of (mean) internal energy,
\cite{igarashi,qian}.

By assuming that $\rho V {v}$  vanishes at integration volume
boundaries (or infinity), we easily get the time rate of Helmholtz
free energy at a constant temperature $T$:
\begin{equation}
{\frac{d}{dt}} \left< \Psi  \right> = -  k_BT \dot{{\cal{Q}}}  - T
\dot{\cal{S}}_G \, . \label{temp}
\end{equation}

By employing Eq.~(\ref{balance1}),  we readily  arrive at
\begin{equation}
 {\frac{d}{dt}} \left< \Psi  \right>  = -  (k_BT ) \left( {\frac{d{\cal{S}}}{dt}}\right)_{in}  = - (m\beta )
 \left< {v}^2\right>  \label{growth}
\end{equation}
which   either identically vanishes (equilibrium)  or remains
negative.

 Thus, Helmholtz free energy either remains constant in time
or decreases as a function of time at  the rate set by the
information entropy "production"   $\dot{\cal{S}}_{in}$. One may
expect that actualy $\langle \Psi \rangle (t)$ drops down to a
finite minimum as $t\rightarrow \infty $.

 However, this feature is a little bit  deceiving. One should be aware that
 a finite minimum  as well may  not exist, which is the case e.g.  for the
  free Brownian.  Also, multiple minima need to be excluded as
  well.

\subsection{Kullback entropy versus Shannon entropy in the
Smoluchowski process}

 In the presence of external forces  the property Eq.~(\ref{growth})  may consistently  quantify
 an asymptotic approach towards a minimum corresponding to an invariant (presumed
to be unique) probability  density of the process. Indeed, by
invoking Eq.~(\ref{eq}) we realize that
\begin{equation}
\rho _*(x)= {\frac{1}{Z}} \exp \left(-{\frac{V(x)}{k_BT}}\right)
\end{equation}
 where $Z=\int \exp(-V(x)/k_BT)\, dx $, sets
the   minimum   of  $\langle \Psi \rangle (t)$ at $\langle \Psi
\rangle _* = \Psi _*= -k_BT \ln  Z$.

Let us take the above $\rho _*(x)$  as a reference density  with
respect to which the divergence of $\rho (x,t)$ is evaluated in
the course of the  pertinent Smoluchowski process. This divergence
is well quantified  by the conditional  Kullback entropy
${\cal{H}}_c(t)$.  Let us notice that
\begin{equation}
{\cal{H}}_c(t) =  -  \int     \rho \, \ln \left({\frac{\rho }{\rho
_*}}\right )\, dx  =  {\cal{S}}(t) - \ln Z - {\frac{\langle
V\rangle }{k_BT}}
\end{equation}

Consequently, in view of Eqs.~(\ref{temp})  and (\ref{balance1}),
we get
\begin{equation}
\dot{\cal{H}}_c = \dot{\cal{S}} + \dot{\cal{Q}} =
(\dot{\cal{S}})_{in} \geq 0
\end{equation}
so   that ${\frac{d}{dt}} \langle \Psi \rangle =  - (k_BT)\,
\dot{\cal{H}}_c$. An approach of $\langle \Psi \rangle (t)$
towards the minimum  proceeds in the very same rate as this of
${\cal{H}}_c(t)$   towards its maximum.

In contrast to $\dot{\cal{H}}_c$ which is non-negative, we have no
growth guarantee for the differential entropy $\dot{\cal{S}}$
whose sign is unspecified. Nonetheless, the balance between the
time  rate of entropy  production/removal   and the power release
into or out of the  environment, is  definitely correct. We have
 $\dot{\cal{S}} \geq - \dot{\cal{Q}}$.

 The  relationship between two different  forms
 of entropy, differential (Shannon)  and conditional (Kullback),
  and their dynamics is thereby set.

An exhaustive discussion of the temporal approach to equilibrium
of the Gibbs and conditional entropies can be found in Refs.
\cite{tyran,tyran1}. Both invertible deterministic and
non-invertible stochastic systems were addresses.

Typically, the conditional entropy either remains constant or
monotonically increases to its maximum at zero. This stays  in
conformity with expectations motivated by the Boltzamnn
$H$-theorem and the second law of thermodynamics.

 To the  contrary, the Gibbs entropy displays a different behavior, even
under the very same approach-to-equilibrium circumstances: it may
monotonically increase or decrease, and as well  display an
oscillatory pattern.  In below we shall  demonstrate that this
potentially strange behavior of the differential entropy gives an
insight into nontrivial power transfer processes in the mean,
whose assessment would not be possible in terms of the conditional
entropy.

\subsection{One-dimensional Ornstein-Uhlenbeck process}

It is quite illuminating to exemplify previous considerations  by
a detailed presentation of the standard one-dimensional
Ornstein-Uhlenbeck process. We denote $b(x)= - \gamma x  $ with
$\gamma >0$.

If  an initial density is chosen in the Gaussian form, with the
mean value $\alpha _0$ and variance $\sigma ^2_0$. the
Fokker-Planck evolution  Eq.~(\ref{Fokker})  preserves the
Gaussian form  of $\rho (x,t)$ while modifying
 the mean value
$\alpha (t) = \alpha _0 \exp(-\gamma t)$ and variance according to
\begin{equation}
 \sigma ^2(t) = \sigma ^2_0 \exp (- 2 \gamma t) + {\frac{D}{\gamma }}[1-\exp (-2\gamma t)] \, .
\end{equation}

Accordingly, since   a unique invariant density has the form $\rho
_* =   \sqrt{\gamma /2\pi D}
 \exp (-\gamma x^2/2D)$ we obtain, \cite{tyran}:
\begin{equation}
{\cal{H}}_c(t) = \exp (-2 \gamma t) {\cal{H}}_c(\rho _0,\rho _*)=
-{\frac{\gamma \alpha _0^2}{2D}}\, \exp (-2\gamma t)
\end{equation}
while in view of our previous considerations, we have
${\cal{S}}(t) =(1/2) \ln [2\pi e \sigma ^2(t)] $ and ${\cal{F}} =
1/\sigma ^2(t)$.  Therefore
\begin{equation}
\dot{\cal{S}}= {\frac{2\gamma  (D - \gamma \sigma _0^2) \exp(-2
\gamma t)} {D -(D - \gamma \sigma _0^2)\exp(-2\gamma t)}}\, .
\end{equation}

We observe that if $\sigma ^2_0 > D/\gamma $, then $\dot{\cal{S}}
<0$, while $\sigma ^2_0 < D/\gamma $ implies $\dot{\cal{S}} > 0$.
In both cases the behavior of the differential   entropy is
monotonic, though its growth or decay do critically rely on  the
choice of $\sigma ^2_0$. Irrespective of $\sigma ^2_0$ the
asymptotic value of   ${\cal{S}}(t)$ as  $t\rightarrow \infty $
reads $(1/2) \ln [2\pi e (D/\gamma )$.

The differential entropy evolution is anti-correlated with  this
of the localization, since
\begin{equation}
\dot{\cal{F}}= -\,  {\frac{\gamma \dot{\cal{S}}} {[D -(D - \gamma
\sigma _0^2)\exp(-2\gamma t)]^2}}\, .
\end{equation}
For all $\sigma _0^2$  the asymptotic value of ${\cal{F}}$  reads
$\gamma /D$.

We have here a direct control of the behavior of the "power
release" expression $\dot{\cal{Q}} = \dot{\cal{H}}_c -
\dot{\cal{S}}$.  Since
\begin{equation}
\dot{\cal{H}}_c= (\gamma ^2\alpha _0^2/D) \exp(-2\gamma t) >0\, ,
\end{equation}
in case of $\dot{\cal{S}} <0$ we encounter
 a continual power supply $\dot{\cal{Q}}> 0$  by the thermal environment.

 In case of $\dot{\cal{S}} >0$ the situation is more complicated. For example, if $\alpha _0 =0$, we
  can easily check that  $\dot{\cal{Q}}  < 0$, i.e. we have the power drainage from the environment  for all $t\in R^+$.
  More generally, the sign of  $\dot{\cal{Q}}$ is
  negative for   $\alpha _0^2< 2(D-\gamma \sigma ^2_0)/\gamma$. If the latter inequality  is reversed, the sign of
  $\dot{\cal{Q}}$ is not uniquely specified  and suffers  a change at a suitable time  instant $t_{change}
  (\alpha _0^2,\sigma _0^2)$.

\subsection{Mean energy and  the dynamics of Fisher information}

By considering   $( - \rho )(x,t)$ and
 $s(x,t)$, such that
$v= {\nabla }s$, as canonically conjugate
 fields,  we can  invoke the variational calculus. Namely, one may
   derive  the continuity (and thus Fokker-Planck) equation together
 with  the Hamilton-Jacobi type equation (whose gradient implies the hydrodynamical
 conservation law Eq.~(\ref{law})):
\begin{equation}
\partial _ts +\frac{1}2 ({\nabla }s)^2 - (\Omega -  Q) = 0
\label{jacobi}\, ,
\end{equation}
 by means of the extremal (least, with fixed end-point variations) action
principle
 involving  the (mean} Lagrangian:
\begin{equation}
{\cal{L}} = - \int \rho\left[ \partial _t s   +  {\frac{1}2}
({\nabla }s)^2  - \left({\frac{{u}^2}2} + \Omega \right) \right]
dx\, .
\end{equation}

The related  Hamiltonian  (which is  the mean energy of the
diffusion process  per unit of mass) reads
\begin{equation}
{\cal{H}} \doteq \int  \rho \cdot \left[ {\frac{1}2}({\nabla }
s)^2  - \left( {\frac{{u}^2}2} + \Omega \right ) \right] \, dx
\label{energy}
\end{equation}
i. e.
$${\cal{H}} = (1/2) (\left< {v}^2\right> - \left< {u}^2\right>)  -
\left<\Omega \right> \, .
$$

We can evaluate an expectation value of Eq. (\ref{jacobi}) which
implies an identity ${\cal{H}} = - \left< \partial _ts \right>$.
By invoking Eq.~(\ref{Helmholtz}),  with the
 time-independent $V$,  we  arrive at
\begin{equation}
 \dot{\Psi } = {\frac{ k_BT}{\rho }}  \nabla (v\rho )
\end{equation}
  whose   expectation value  $\langle \dot{\Psi } \rangle$, in view of
   $v\rho =0$  at   the integration volume boundaries,  identically  vanishes.
   Since   $v= - (1/m\beta ) \nabla \Psi $,  we  define
   \begin{equation}
   s(x,t)\doteq (1/m\beta ) \Psi (x,t)  \Longrightarrow  \left< \partial _ts \right>=0
   \end{equation}
so that ${\cal{H}} \equiv 0$ identically.

 We have  thus  arrived at the  following     interplay between the mean energy
 and the information entropy "production" rate:
  \begin{equation}
{\frac{D}2} \left( {\frac{dS}{dt}}\right)_{in} =
{\frac{1}2}\langle v^2\rangle =
  \int  \rho \, \left(
{{\overrightarrow{u}}^2\over 2} + \Omega \right) \, dx  \geq 0  \,
,
\end{equation}
generally valid for Smoluchowski processes  with non-vanishing
diffusion currents.

By recalling the notion of the Fisher information
Eq.~(\ref{Fisher1}) and setting ${\cal{F}} \doteq  D^2
{\cal{F}}_{\alpha }$,  we can rewrite the above formula as
follows:
\begin{equation}
{\cal{F}}  =  \langle v^2 \rangle - 2 \langle \Omega \rangle \geq
0   \label{interplay}
\end{equation}
 where  $
{\cal{F}}/2 = - \langle Q \rangle >0$  holds true for probability
densities with  finite mean and variance.

We may evaluate directly the uncertainty dynamics of the
Smoluchowski process, by recalling that the Fisher information
${\cal{F}}/2$ is the localization measure, which for probability
densities with   finite mean value and
 variance $\sigma ^2$ is  bounded from below by $1/\sigma ^2$, see e.g. Section III.

Namely, by exploiting the  hydrodynamical  conservation laws
Eq.~(\ref{law})  for the Smoluchowski process   we get:
\begin{equation}
\partial _t (\rho {v}^2) = - {\nabla}\cdot
 [(\rho  {v}^3)] - 2\rho {v}\cdot
 {\nabla }(Q - \Omega) \, .
 \end{equation}
We assume to have  secured conditions allowing to take a
derivative under an indefinite integral, and  take for granted
that  of $\rho  {v}^3$  vanishes
 at the integration volume  boundaries. This implies
 the following expression for the time derivative of $\left< {v}^2\right>$:
\begin{equation}
 {\frac{d}{dt}}{\left< {v}^2\right>} = 2 \left< {v}\cdot {\nabla }
(\Omega - Q) \right> \, .
\end{equation}

Proceeding in the same vein, in view of $\dot{\Omega }$,  we find
that
\begin{equation}
{\frac{d}{dt}} \langle \Omega \rangle =  \langle v \cdot \nabla
\Omega \rangle
\end{equation}
and so the equation of motion for ${\cal{F}}$ follows:
\begin{equation}
{\frac{d}{dt}} {\cal{F}} = {\frac{d}{dt}} [ \langle v^2\rangle -
2\langle \Omega \rangle  ]=
   -  2 \langle v\cdot \nabla Q \rangle \, .
\label{Fishdynamics}
\end{equation}

Since  we have $\nabla Q = \nabla P/\rho $ where $P=D^2\rho\,
\Delta \ln \rho $, the previous equation takes the form
$\dot{\cal{F}} =  - \int \rho v \nabla Q dx =  - \int v \nabla P
dx$, which is an analog of the familiar expression for the power
release ($dE/dt = F\cdot v$, with $F=-\nabla V$) in classical
mechanics;
 this to be compared with our  previous discussion of the "heat dissipation"
 term Eq.~(\ref{heat1}).

{\bf Remark 11:} As  indicates our previous example of the
Ornstein-Uhlenbeck  process in one dimension, there is nothing
obvious
 to say about the  growth or decay of various quantities involved.  In this  particular case, we have
e.g.  $\langle v^2\rangle (t) = (D/2) \dot{\cal{H}}_c = t(\gamma
^2 \alpha _0^2/D) \exp(-2 \gamma t)$, hence an asymptotic value
$0$, while $\langle u^2 \rangle (t) = (D/2) {\cal{F}}(t)
\rightarrow  \gamma /D$. Accordingly, we have $\langle \Omega
\rangle (t) \rightarrow - \gamma /2D$.

\section{Differential entropy  dynamics in quantum theory}

\subsection{Balance equations}

In the discussion of Smoluchowski diffusions, our  major reference
point  was the conventional
 Fokker-Planck equation (\ref{Fokker})  for a probability density  supporting a
  Markovian diffusion process. The (time-independent)  drift function  $b$
  was assumed to be known a priori
 (e.g. the conservative  external forces were  established on  phenomenological or model construction
 grounds), while  the initial and/or boundary data for the probability density of the
 process  could be chosen (to a high degree) arbitrarily.

Under such  "normal"  circumstances, the  hydrodynamical
conservation laws (\ref{law}) come   out as  a direct consequence
of the Fokker-Planck equation. Also, the functional expression for
$\Omega $ of the form (\ref{Omega}) is basically known to arise if
one attempts to replace an elliptic diffusion  operator by a
Hermitian (and possibly self-adjoint) one,
\cite{risken,qian1,gar}.

We shall depart from the standard Brownian motion setting to more
general  Markovian diffusion-type  processes which, while still
respecting
 the Fokker-Planck equation, admit general time-dependent  forward drifts.
 In fact,  we invoke  at this  point a well defined  stochastic counterpart of the
 Schr\"{o}dinger picture  quantum dynamics  of wave
 packets, \cite{nelson,carlen,gar,gar1,eberle,qian1,carlen1},  where  the notion of
 differential entropy and its dynamics   finds   a proper
 place. The dynamics of  quantal probability densities is here resolved
 in terms of diffusion-type processes.

 Let us assume to have chosen  an arbitrary continuous  (it is useful, if bounded from below)
 function ${\cal{V}}= {\cal{V}}(\overrightarrow{x},t)$ with dimensions of
 energy.
we consider the Schr\"{o}dinger equation (set $D=\hbar /2m$)  in
the form
\begin{equation}
i\partial _t  \psi  = - D  \Delta  \psi   +
{\frac{{\cal{V}}}{2mD}} \label{Schroedinger} \psi \, .
\end{equation}

The Madelung decomposition $\psi = \rho ^{1/2} \exp(is)$  with the
phase function $s=s(x,t)$
 defining $v=\nabla s $ is known to imply two coupled equations: the standard continuity equation
 $\partial _t \rho = - \nabla (v \rho )$ and the Hamilton-Jacobi-type equation
\begin{equation}
\partial _ts +\frac{1}2 ({\nabla }s)^2 + (\Omega -  Q) = 0 \label{jacobi1}
\end{equation}
where $\Omega \doteq {\cal{V}}/m$  and the functional form  of $Q$
coincides with this introduced previously in
Eq.~(\ref{potential1}). Notice  a "minor" sign change in
Eq.~(\ref{jacobi1}) in comparison  with Eq.~(\ref{jacobi}).

Those two equations   form a coupled system, whose solutions
describe a Markovian diffusion-type process: the probability
density is propagated by a   Fokker-Planck dynamics of the form
Eq.~(\ref{Fokker})  with the drift $b=v-u$ where $u=D\nabla \ln
\rho $ is an osmotic velocity field.

We can mimic the  calculus of variations steps of the previous
section, so arriving at the  Hamiltonian  (actually, the mean
energy of the quantum motion per unit of mass):
\begin{equation}
{\cal{H}} \doteq \int  \rho \cdot \left[ {\frac{1}2}({\nabla }
s)^2  + \left( {\frac{{u}^2}2} + \Omega \right ) \right] \, dx \,
,
\end{equation}
to be compared with Eq.~(\ref{energy}).  There holds
\begin{equation}
{\cal{H}} = (1/2) [\left< {v}^2\right> + \left< {u}^2\right>]  +
\left<\Omega \right>  =  - \left< \partial _t s\right>    \, .
\end{equation}

Of particular interest (due to its relative simplicity) is the
case of time-independent ${\cal{V}}$, when
\begin{equation}
{\cal{H}} = - \left< \partial _t s\right> \doteq {\cal{E}} = const
\end{equation}
is  known to be a conserved finite
 quantity, which is not necessarily positive.  Since  generally ${\cal{H}}
   \neq 0$, we deal here  with so-called
finite energy diffusion-type  processes, \cite{carlen,eberle}. The
corresponding Fokker-Planck equation propagates  a probability
density  $|\psi |^2 = \rho $,
 whose  differential  entropy ${\cal{S}}$ may quite nontrivially evolve in time.

Keeping intact the previous derivation procedures  for
$(\dot{\cal{S}})_{in}$   (while  assuming the validity of
mathematical restrictions upon the behavior of integrands), we
encounter the information entropy balance equations in  their
general form disclosed in Eqs.~(\ref{balance})-(\ref{balance1}).
The related differential
 entropy "production" rate reads:
\begin{equation}
 (\dot{\cal{S}})_{in}  =
 {\frac{2}D} \left[ {\cal{E}}  - \left({\frac{1}2} {\cal{F}} +
  \left<\Omega \right>\right) \right]  \geq 0 \, , \label{interplay1} \, .
 \end{equation}

 We recall that ${\frac{1}2} {\cal{F}} = - \langle Q \rangle >0 $ which implies
 $ {\cal{E}} - \langle \Omega \rangle \geq {\frac{1}2}  {\cal{F}} > 0$. Therefore,
   the localization measure
$\cal{F}$  has a definite upper bound:  the  pertinent wave packet
cannot
 be localized too sharply.

We notice that  the localization  (Fisher) measure
\begin{equation}
 {\cal{F}} = 2({\cal{E}} -  \langle  \Omega \rangle ) -
 \langle v^2 \rangle
\end{equation}
in general evolves in time.
  Here  ${\cal{E}}$ is a constant and $\dot{\Omega }=0$.

By invoking the hydrodynamical conservation laws, we find out that
 the dynamics of  Fisher information  follows an equation:
\begin{equation}
{\frac{d{\cal{F}}}{dt}} =  + 2 \langle v \nabla Q \rangle
\end{equation}
and that there holds
\begin{equation}
{\frac{1}2} \dot{\cal{F}}  =  -   {\frac{d}{dt}} [{\frac{1}2}
\langle v^2\rangle + \langle \Omega \rangle ]
\label{Fishdynamics1}
\end{equation}
which is to be compared (notice the opposite sign of the
right-hand expression) with the result we have obtained  for
Smoluchowski processes.

 Obviously, now we have $\dot{\cal{F}} =
+ \int v \nabla P dx$, with  the same functional form for $P$ as
before. We
 interpret $\dot{\cal{F}}$ as the measure of  power
transfer in the course of which the (de)localization "feeds" the
diffusion current and  \it in reverse. \rm
 Here, we encounter  a negative feedback between the
localization and the proper energy of motion which keeps intact an
overall mean energy    ${\cal{H}} = {\cal{E}}$ of the quantum
motion. See e.g. also \cite{gar}.

  In case of $v=0$, we have
 ${\cal{E}} = {\frac{1}2} {\cal{F}} +
  \langle \Omega \rangle $
and no entropy "production" nor dynamics of uncertainty.
 There holds   $\dot{\cal{S}} =0$  and  we deal with  time-reversible
stationary diffusion  processes and  their invariant probability
densities $\rho (x)$, \cite{qian1,eberle}.

{\bf  Remark 12:}  Let us indicate  that  the phase  function
 $s(x,t)$  shows up certain (remnant)  features of the Helmholtz $\Psi $ and
 $\langle \Psi \rangle $.  This behavior is not unexpected, since
 e.g. the ground state  densities (and  other invariant  densities of
 stationary states) are directly related to time-reversible stationary
 diffusion-type processes of Refs.~\cite{eberle,qian1}.
We have $- \langle \partial _t s\rangle = {\cal{E}}$. In view of
$v=\nabla s$ and assumed vanishing of $s\rho v$ at  the
integration volume boundaries, we get:
\begin{equation}
{\frac{d}{dt}} \langle s \rangle =  \langle v^2 \rangle -
{\cal{E}}
\end{equation}
The previously mentioned case of  \it  no  \rm  entropy
"production" refers to $v=0$ and thus $s=s_0 - {\cal{E}}\cdot t$.

We  recall that  the corresponding derivation of
Eq.~(\ref{growth}) has been carried  out for $v=- (1/m\beta
)\nabla \Psi $,  with  $\langle \dot{\Psi }\rangle =0$). Hence, as
close as possible link with the present discussion is obtained if
we re-define $s$ into $s_{\Psi } \doteq -s $. Then we have
\begin{equation}
{\frac{d}{dt}} \langle s_{\Psi } \rangle =  {\cal{E}}  -   \langle
v^2 \rangle \, .
\end{equation}
For stationary quantum states,  when $v=0$ identically,  we get
${\frac{d}{dt}} \langle s_{\Psi } \rangle =  {\cal{E}}$, in
contrast to the standard Fokker-Planck case of ${\frac{d}{dt}}
\langle \Psi  \rangle =0$.

Interestingly enough, we can write the generalized Hamilton-Jacobi
equation, while  specified to  the  $v=$ regime, with respect to
$s_{\Psi }$. Indeed, there holds   $ \partial _t s_{\Psi } =
\Omega - Q$, in close affinity with Eq.~(\ref{jacobi}) in the same
regime.

\subsection{Differential entropy dynamics exemplified}

\subsubsection{Free evolution}

Let us consider  the probability density in one space dimension:
\begin{equation}
\rho (x,t) = {\frac{\alpha }{[\pi (\alpha ^4 + 4D^2t^2)]^{1/2}}}
\exp \left( - {\frac{x^2\alpha ^2}{\alpha ^4 + 4D^2t^2}} \right)
\end{equation}
and the phase  function
\begin{equation}
s(x,t) = {\frac{2D^2x^2t}{\alpha ^4 + 4D^2t^2}} - D^2 \arctan
\left( - {\frac{2Dt}{\alpha ^2}}\right)
\end{equation}
which determine  a free wave packet solution of  equations
(\ref{Schroedinger}) and (\ref{jacobi1}), i.e.  obtained  for the
case of ${\cal{V}} \equiv 0$ with the initial data $\psi (x,0)=
(\pi \alpha ^2)^{-1/4} \exp(- x^2/2 \alpha ^2)$.

We have:
\begin{equation}
b(x,t)= v(x,t) + u(x,t) = {\frac{2D(2Dt - \alpha ^2)x}{\alpha ^4 +
4D^2t^2}}
\end{equation}
and the  the Fokker-Planck equation  with the  forward drift
$b(x,t)$ is solved by  the above  $\rho $.

In the present case, the differential  entropy reads:
\begin{equation}
{\cal{S}}(t) = {\frac{1}{2}} \ln \left[ 2 \pi e \left<
X^2\right>(t)\right]
\end{equation}
where $\left< X^2\right> \doteq \int x^2\rho dx = (\alpha ^4 +
4D^2t^2)/2\alpha ^2$. Its time rate  $D {\dot{\cal{S}}} = \langle
v^2 \rangle  - \langle b \cdot v \rangle $ equals:
\begin{equation}
 D \frac{d{\cal{S}}}{dt} = {\frac{4D^3t}{\alpha ^4 + 4D^2t^2}} \geq  0
\end{equation}
for $t\geq 0$. Its large time asymptotic is $D/t$.

Furthermore, we have
\begin{equation}
 D (\dot{\cal{S}})_{in} =   \left< v^2\right> =
\frac{8D^4t^2}{\alpha ^2(\alpha ^4  + 4D^2 t^2)}
\end{equation}
with the obvious large time asymptotic value $2D^2/\alpha ^2$: the
differential entropy  production remains untamed for all times.

Due to  $\langle u^2 \rangle = (2D^2 \alpha ^2)/(\alpha ^4 +
4D^2t^2)$ there holds
\begin{equation}
{\cal{E}}=  {\frac{1}2} (\langle v^2\rangle + \langle u^2 \rangle
) = {\frac{D^2}{\alpha ^2}} \, .
\end{equation}
Accordingly,  the quantum mechanical analog of the  entropy
(rather than heat) "dissipation" term $- D\cdot {\cal{Q}}$   in
the quantum case reads
\begin{equation}
 -  \langle b \cdot v  \rangle  = \frac{4D^3t( \alpha ^2 -2Dt)}{\alpha ^2 (\alpha ^4 + 4D^2t^2)}
\end{equation}
and while  taking negative values for $t< \alpha ^2/2D$, it  turns
out to be positive for larger times. Formally speaking, after a
short   entropy "dissipation" period  we pass to the entropy
"absorption"
 regime which in view of its $D/t$ asymptotic,   for large times  definitely   dominates
 $D (\dot{\cal{S}})_{in} \sim
 2D^2/\alpha ^2$.

Those differential entropy balance features do parallel a
continual growth of the mean kinetic energy $(1/2)\langle v^2
\rangle $  from an initial value $0$ towards its asymptotic value
 $D^2/\alpha ^2 = {\cal{E}}$.  Note that the negative  feedback is here displayed
 by the the behavior  of $\langle u^2  \rangle $ which  drops down from the
  initial  value $2D^2/\alpha ^2$ towards  $0$.
It is also instructive to notice that in the present case
${\cal{F}}(t) = D^2/\langle X^2\rangle (t)$. We can readily check
that $\dot{\cal{F}} = d\langle u^2 \rangle /dt =   - d\langle
v^2\rangle /dt$.

\subsubsection{Steady state}

We choose the probability density in the form:
\begin{equation}
\rho (x,t) = \left( {\frac{\omega } {2\pi D}}\right) ^{1/2} \exp
\left[ - {\frac{\omega }{2D}}\, \left( x - q(t)\right)^2 \right]
\end{equation}
where the classical harmonic dynamics with particle mass $m$ and
 frequency $\omega $ is involved such that
$q(t)= q_0 \cos (\omega t) + (p_0/m\omega ) \sin (\omega t)$ and
$p(t) = p_0\cos (\omega t) - m\omega q_0 \sin (\omega t)$.

One can easily verify that (9),   and (40) hold true identically
once we set ${\cal{V}}=  {\frac{1}2}\omega ^2 x^2$  and  consider:
\begin{equation}
  s(x,t) = (1/2m)\left[ xp(t) -
(1/2)p(t)q(t) - mD\omega t\right] \, .
\end{equation}

A  forward drift takes the form:
\begin{equation}
b(x,t) = {\frac{1}m} p(t) - \omega \left(x-q(t)\right)
\end{equation}
and  the above  $\rho $ solves  the  corresponding  Fokker-Planck
equation.

The differential entropy is a constant equal  ${\cal{S}}= (1/2)
\ln (2\pi e D/\omega )$.  Although trivially $d{\cal{S}}/dt =0$,
all previous arguments can be verified.

For  example, we have $v=\nabla  s= p(t)/2m$ and therefore an
oscillating entropy "production" term $D (\dot{\cal{S}})_{in} =
p^2(t)/4m^2$  which   is balanced
 by  an oscillating "dissipative"  counter-term to yield  $\dot{\cal{S}}$. Suitable
 expressions for $\langle s\rangle $ and $\langle \partial _ts \rangle $ easily follow.

Concerning the Fisher measure, we have obviously ${\cal{F}} =
\omega /D$ which is a constant of motion.

\subsubsection{Squeezed state}

Let us  consider \cite{majernik} the squeezed  wave function of
the  harmonic oscillator. We adopt the re-scaled units $\hbar
=\omega =m=1$, hence also $D=1$.  The solution of the Schr\"{o}d
inger equation $i\partial _t\psi = -(1/2)  \Delta \psi  + (x^2/2)
\psi$ with the initial data $\psi (x,0) = (\gamma ^2 \pi )^{-1/4}
\exp(-x^2/2\gamma ^2)$  and $\gamma \in (0,\infty )$,  is defined
in terms of  the probability density :
\begin{equation}
\rho (x,t) = {\frac{1}{(2\pi )^{1/2}\sigma(t)}} \exp \left( -
{\frac{x^2}{2\sigma ^2(t)}} \right)
\end{equation}
where
\begin{equation}
2\sigma ^2(t) = {\frac{1}{\gamma ^2}} \sin t + \gamma ^2 \cos^2t
\end{equation}
and the phase function
\begin{equation}
s(x,t) = {\frac{t}2}  + {\frac{\pi }4} +  \arctan (\gamma ^2 \cot
t)  + {\frac{(1/\gamma ^2 - \gamma ^2)\sin 2t}{8\sigma ^2(t)}} x^2
\, .
\end{equation}

 Now,  the differential  entropy   ${\cal{S}}= (1/2) \ln
[2\pi e \sigma ^2(t)]$  displays a periodic behavior in time,
whose level of complexity depends on the particular value of the
squeezing parameter $\gamma $. The previously mentioned  negative
feedback is here manifested  through  (counter)oscillations of the
localization, this in conformity with the dynamics of $\sigma
^2(t)$ and the corresponding oscillating dynamics of the Fisher
measure ${\cal{F}}= 1/\sigma ^2(t)$. .

See e.g. also  \cite{majernik} for a pictorial analysis and   an
instructive computer assisted discussion
 of the  Schr\"{o}dinger cat state (superposition of the harmonic oscillator  coherent states with the same
 amplitude but with opposite phases), with the  time evolution of the  corresponding differential entropy.

\subsubsection{Stationary states}

In contrast to generic applications of the standard Fokker-Planck
equation, where one takes for granted that there is a unique
positive stationary probability density, the situation looks
otherwise if we admit the Schr\"{o}dinger equation as a primary
dynamical rule  for the  evolution of (inferred)  probability
densities. For a chosen potential, all available stationary
quantum  states may  serve the purpose, since then we have
nonnegative (zeroes are now admitted) $\rho _*(x)$, and $v(x)=0$
identically  (we stay in one spatial
 dimension).

The standard harmonic oscillator may serve  as an instructive
example.  One may e.g. consult Fig. $3$ in \cite{yanez} to check
the behavior of both position and momentum differential entropies,
and their  sum, depending on the energy eigenvalue. All these
stationary state  values grow monotonically   with $n=1,2,...,60$,
\cite{yanez} and follow the pattern in the asymptotic regime
$n\equiv 500$, \cite{majernik}.

For convenience we shall refer to  the  Schr\"{o}dinger eigenvalue
problem with  scaled away physical units. We consider ( compare
e.g. Eq.~(\ref{Schroedinger}) with $D \rightarrow 1/2$)
\begin{equation}
[-{\frac{1}2} \Delta + {\frac{x^2}2}] \sqrt{\rho _*}  =
(n+{\frac{1}2})  \sqrt{\rho _*} \, .
\end{equation}
In terms of a suitable Hamilton-Jacobi type  equation we can
address the same problem by seeking
 solutions of an equation
\begin{equation}
 n+ 1/2  = \Omega - Q \label{hj}
\end{equation}
  with respect to $\sqrt{\rho _*}$,   provided we set $\Omega = x^2/2$, define   $u = \nabla \ln \sqrt{\rho _*}$
  and demand   that $Q= u^2/2 + (1/2)\nabla \cdot u$.

For the harmonic oscillator problem, we can refer to standard
textbooks.  For each value of $n$ we recover a corresponding
unique stationary density:  $\sqrt{\rho _*}\rightarrow \rho
_n^{1/2}$ with $n=0,1,2,...$). We   have:
\begin{equation}
\rho _n^{1/2}(x) = {\frac{1}{(2^n n! \sqrt{\pi })^{1/2}}} \, \exp
\left( - {\frac{x^2}2}\right) \, H_n(x)
\end{equation}
where  $H_n(x)$ stands for the $n$-th Hermite polynomial: $H_0=1$,
$H_1=2x$, $H_2= 2(2x^2 -1)$, $H_3 = 4x(2x^2-3)$, and so on.

We immediately infer e.g.  $b_0=-x  \rightarrow  Q = x^2/2  -
1/2$, \, next $b_1= (1/x) - x \rightarrow  Q= x^2/2 - 3/2$, and
$b_2= [4x/(2x^2 - 1)] - x  \rightarrow  Q = x^2/2 - 5/2$, plus
$b_3= [(1/x) + 4x/(2x^2 -3)] \rightarrow Q = x^2-7/2$,  that is to
be  continued for $n>3$. Therefore  Eq.~(\ref{hj}) is here a
trivial identity.

Obviously, except for the ground state which is strictly positive,
all remaining stationary states are nonnegative.

An open problem, first generally addressed in \cite{fortet}, see
also \cite{blanch},
 is to implement a continuous dynamical process for which any of induced stationary densities  may serve
 as an invariant asymptotic  one.  An obvious (Ornstein-Uhlenbeck) solution is known for the ground state
  density.  Its ample discussion (albeit without mentioning the quantum connotation) has been
   given in Section IV.

\section{Outlook}

 One may raise an issue of what are the entropy functionals  good for.
 In non-equilibrium statistical  mechanics of gases one invokes them to solve concrete
 physical problems: for example to  address the second law of thermodynamics and the
related Boltzmann $H$-theorem from a probabilistic  point of view.

 We are strongly   motivated by a  number of reported problems with  the rigorous
formulation of effects of noise on entropy evolution and a
justification of the \it  entropy growth  \rm paradigm for model
systems, \cite{cercignani,huang,hasegawa,tyran,jaynes,jaynes1} and
a long list of mathematically oriented papers on the  large time
asymptotic of stochastic  systems,
 \cite{voigt}-\cite{toscani2}.
Therefore, the major issue  addressed in the  present  paper is
that of quantifying the dynamics  of probability densities in
terms of proper entropy functionals. The formalism was designed to
encompass standard diffusion processes of non-equilibrium
statistical physics and the Schr\"{o}dinger picture implemented
dynamics of probability densities related to pure quantum states
in $L^2(R)$. In the latter case  an approach towards equilibrium
has not been expected to occur at all.

To  this end the behavior of Shannon and Kullback-Leibler
entropies in time has been investigated in classical and quantum
mechanical  contexts. The utility of a particular form of entropy
for a given dynamical  model appears to be basically purpose
dependent. The use of the Kullback-Leibler  entropy encounters
limitations which are not shared by the differential entropy, when
the dynamical process is rapid. Alternatively, if  one is
interested in its short-time   features.

 On the contrary, it is the  conditional  Kullback-Leibler entropy which
 is often invoked  in rigorous formulations of the Boltzmann  $H$-theorem
 under almost-equilibrium conditions
and its analogues for stochastic systems.  The large time
asymptotics of  solutions of   Fokker-Planck equations,  if
analyzed in terms of this  entropy,  gives reliable results.

  However, our  analysis of Smoluchowski diffusion processes and  of  the  exemplary
   Ornstein-Uhlenbeck process,   demonstrates  that  a deeper insight into   the underlying non-equilibrium
     physical phenomena (the  inherent power  transfer processes) is  available  only  in terms of the
     Shannon entropy and its time rate of change. This  insight is inaccessible  in terms  of the
         Kullback-Leibler entropy .

The differential entropy   needs not to  increase, even in case of
plainly irreversible dynamics. The  monotonic growth  in time of
the conditional Kullback entropy (when applicable), not
necessarily should be related to the "dynamical origins of
 the increasing entropy", \cite{mackey,tyran1}.  We would rather
 tell that the conditonal entropy is well suited to stay in correspondence with  the lore
 of the second law of thermodynamics, since by construction its  time behavior is
 monotonic, if one  quantifies  an  asymptotic  approach towards  a stationary
 density.

   In case of Smoluchowski processes, the  time rate of the  conditional Kullback
 entropy was  found to coincide with the corresponding differential (Shannon) entropy "production"
 rate. The differential entropy itself  needs not to grow and may as well change its dynamical
 regime from growth to decay and in reverse, even with the  entropy "production" involved.

 Balance equations for the differential entropy and the  Fisher information measure  involve a nontrivial
 power transfer.  In case of Smoluchowski processes this power release can be easily
  attributed to  the entropy removal  from the system  or the  entropy absorption from (drainage)
   the thermostat.  In the quantum mechanical regime, the  inherent  power transfer   is related to
   metamorphoses of various forms  of mean energy   among themselves  and needs not the notion of
   external to the system thermostat.

Apart from the above observations, we have provided a
comprehensive review of  varied appearances of the differential
entropy in the existing literature on  both classical and quantum
dynamical systems. As a byproduct of the general discussion, we
have described  its  specific  quantum  manifestations, in the
specific (pure quantum states)  regime  where the traditional von Neumann entropy  is
 of no much  use.\\

{\bf  Acknowledgement:} The paper has been supported by the Polish
Ministry of Scientific Research and Information Technology under
the  (solicited) grant No PBZ-MIN-008/P03/2003.  I would like to
thank Professor Robert Alicki for help in quest for some hardly
accessible  references. \\

{\bf Dedication:} This paper is dedicated  to Professor Rafael
Sorkin on the occasion of his 60th birthday, with friendly
admiration.

\noindent \copyright 2005 by MDPI (http://www.mdpi.org).
Reproduction for noncommercial purposes permitted.
\end{document}